\documentclass[nofootinbib,amsmath,prd,aps,superscriptaddress,tightenlines,12pt]{revtex4}
\usepackage[hypertex]{hyperref}
\usepackage{graphicx}
\usepackage{epstopdf}
\usepackage{bm}
\usepackage{epsfig}
\usepackage{graphics}
\usepackage{xspace}

\def\nslash{n\!\!\!\slash}

\def\bnslash{\bar n\!\!\!\slash}

\def\OMIT#1{}

\newcommand{\nn}{\nonumber}

\newcommand{\bn}{{\bar n}}
\newcommand{\bea}{\begin{eqnarray}}
\newcommand{\eea}{\end{eqnarray}}

\newcommand{\be}{\begin{equation}}
\newcommand{\ee}{\end{equation}}

\begin{document}
\setlength\baselineskip{17pt}



\title{\bf Transverse Momentum Distributions from Effective Field Theory with Numerical Results }


\author{Sonny Mantry}
\email[]{mantry147@gmail.com}
\author{Frank Petriello}
\email[]{frankjp@physics.wisc.edu}
\affiliation{University of Wisconsin, Madison, WI, 53706}



\newpage
\begin{abstract}
  \vspace*{0.3cm}
  
We derive a factorization theorem for the differential distributions of electroweak gauge bosons in Drell-Yan processes, valid at low transverse momentum, using the Soft-Collinear Effective Theory (SCET).  We present the next-to-leading logarithmic (NLL) transverse momentum distribution for the Z-boson and find good agreement  with Tevatron data collected by the CDF and D0 collaborations. We also give  predictions for the Higgs boson differential distributions at NLL based on a factorization theorem derived in earlier work.  We derive formulae for all quantities needed to study low transverse momentum production of color-neutral particles within the effective theory.  This effective field theory approach is free of Landau poles and can be formulated entirely in momentum space. Consequently, our results are free of  the Landau-pole prescriptions necessary in the standard approach.  
\end{abstract}

\maketitle

\newpage
\tableofcontents
\section{Introduction}

The study of low transverse momentum ($p_T$) production of electroweak gauge and Higgs bosons plays an important role in many studies within and beyond the Standard Model (SM).  The measurement of the $W$-boson mass at hadron colliders requires an understanding of $W$ production at low $p_T$~\cite{Baur:2003jy}.  The search for a Higgs boson decaying into two $W$ bosons requires a jet veto to remove the large $t\bar{t}$ background~\cite{Dittmar:1996ss}.  This cut effectively restricts the Higgs boson to the low-$p_T$ region.  In the kinematic limit of low transverse momentum, large logarithms of the form $\text{ln}(M/p_T)$, where $M$ denotes the mass of $\gamma^{*},W,Z,h$, spoil the perturbative expansion 
based on the strong coupling constant $\alpha_S$.  The logarithms must be resummed to all orders to obtain an accurate prediction.  This 
resummation has been extensively studied in the literature~\cite{Dokshitzer:1978yd,Parisi:1979se,Curci:1979bg,Collins:1981uk,Collins:1984kg,Kauffman:1991jt,Yuan:1991we,Ellis:1997ii,Berger:2002ut,Kulesza:2002rh,Kulesza:2003wn,Bozzi:2003jy,Bozzi:2005wk, Bozzi:2010xn, Gao:2005iu, Idilbi:2005er}.  The standard approach utilizes a Fourier transform from momentum space to impact-parameter space to decouple emissions of multiple gluons while maintaining momentum conservation~\cite{Parisi:1979se}.  This introduces a Landau pole arising from evaluating the strong coupling $\alpha_s(1/b_{\perp}^2)$ for large impact parameters, $b_{\perp} \to \infty$.  The Landau pole must be dealt with for any transverse momentum, even for $p_T \gg  \Lambda_{QCD}$.  The resummed exponent in $b_{\perp}$-space also does not vanish when $p_T$ becomes large, which potentially introduces numerical instabilities in the matching to fixed-order QCD at high transverse momenta.  An alternative approach which addresses these limitations is worth pursuing.

In a previous paper we performed an analysis of transverse-momentum resummation for the example of Higgs-boson~\cite{Mantry:2009qz} production using the Soft-Collinear Effective Theory (SCET)~\cite{Bauer:2000yr, Bauer:2001yt, Bauer:2002nz}.  We derived a factorization theorem for the differential distributions of the Higgs boson.  The resummation of large logarithms in this approach is performed using renormalization-group (RG) evolution in the effective theory.  We introduced several new functions which encode the emission of soft and collinear gluons within the effective theory, which we labeled impact-parameter beam functions (iBFs) and inverse soft functions (iSFs).  Our SCET formulation has several advantages over the standard approach.  The matching from full QCD onto SCET$_{p_T}$ at the hard scale $M$ followed by running to lower scales is done in momentum-space, not in impact-parameter space as in the standard method.  This has the effect of stopping the RG evolution at the scale $\mu_T \sim p_T$,  the natural scale of SCET$_{p_T}$ describing the dynamics of soft and collinear $p_T$ emissions. The factorization theorem can be written entirely in momentum space which is equivalent to the statement that the integral over $b_\perp$ can be performed analytically without running into the Landau pole. As a result, for perturbative values of $p_T$, the transverse momentum distribution is predicted entirely in terms of perturbative functions and the standard initial-state PDFs. Only for non-perturbative values of  $p_T$, one obtains a new non-perturbative function in SCET$_{p_T}$ which is field theoretically well defined and has a computable anomalous dimension.  This is in contrast to the standard approach where a prescription is introduced to avoid the Landau pole even for perturbative values of $p_T$.  Since an analytic integration over impact parameters is possible, we avoid instabilities that occur in the standard approach  in the matching of the resummed exponent to the fixed order result needed in the region of high $p_T$.
 
Our goal in this manuscript is to extend our previous work in several ways.  We formulate the factorization theorem for low $p_T$ production in SCET$_{p_T}$ to account for electroweak gauge boson production in addition to Higgs production.  We present one-loop expressions for all iBFs and iSFs needed to study resummation to the next-to-leading logarithmic (NLL) order of accuracy.  The two-loop results for the iBFs and the iSF are required for a complete resummation to next-to-next-to-leading logarithmic (NNLL) accuracy.  We compare the SCET formulation to the standard Collins-Soper-Sterman (CSS)~\cite{Collins:1984kg} approach to transverse-momentum resummation.  We present numerical results for the Higgs and $Z$-boson $p_T$ distributions at NLL.  We compare the effective theory predictions for $Z$ production with the experimental measurements at the Tevatron~\cite{Abbott:1999yd,Affolder:1999jh}, and find very good agreement with the data.  This demonstrates that our SCET-based approach provides a promising alternate way of studying transverse momentum distributions at hadron colliders.  We outline future directions for the study of transverse momentum resummation within the effective field theory framework. 

Our paper is organized as follows.  In Section~\ref{sec:review} we pedagogically review the factorization theorem derived in our previous work~\cite{Mantry:2009qz} and present its extension to electroweak gauge boson production.  The details of this extension are presented in appendix \ref{appex-1}.  All analytic results for the matching coefficients, iBFs and iSFs required for phenomenology to NLL and partial NNLL accuracy are presented in Section~\ref{sec:fixedorder}.  The structure of the RG running in the effective theory, which resums large logarithms of the form $\text{ln}\, (M/p_T)$, is discussed in Section~\ref{sec:running}.  Simple analytic expressions for the resummed cross sections valid through NLL are shown in Section~\ref{sec:analytic}.  We discuss the relationship between the various quantities appearing in the SCET approach with those appearing in the CSS formulation in section~\ref{sec:css}, and show the consistency of the methods through NLL.  We discuss what further work must be done to establish the relationship to higher orders.  Numerical results for Higgs production and $Z$ boson production are shown in Section~\ref{sec:numbers}, and the agreement with the Tevatron data for $Z$ production is demonstrated.  Finally, we conclude in Section~\ref{sec:conc}.

\section{Review of the factorization theorem}
\label{sec:review}

We begin by summarizing the content and derivation of our previously-studied factorization theorem~\cite{Mantry:2009qz}, and present its extension to the case of electroweak gauge boson production.  The details of this extension are presented in appendix \ref{appex-1}.  The derivation and result of our factorization analysis are shown schematically below:
\begin{eqnarray}
\label{fac-schem}
\frac{d^2 \sigma}{dp_T^2 dY} &\sim& \int PS \, |{\cal M}_{QCD}|^2 \\ \nonumber
&\downarrow& \text{(match QCD to }\text{SCET}_{p_T})\\ \nonumber
&\sim& \int PS \, |C \otimes \langle {\cal O}_{SCET} \rangle |^2 \\ \nonumber
&\downarrow& \text{(SCET soft-collinear decoupling)}\\ \nonumber
&\sim& H \otimes B_n \otimes  B_{\bar{n}} \otimes S \\ \nonumber
&\downarrow& \text{(zero-bin and soft subtraction equivalence)}\\ \nonumber
&\sim& H \otimes  \tilde{B}_n  \otimes \tilde{B}_{\bar{n}}  \otimes S^{-1} \\ \nonumber
&\downarrow& \text{(match }\text{SCET}_{p_T}\text{ to }\text{SCET}_{\Lambda_{QCD}})\\ \nonumber
&\sim& H \otimes \underbrace{\left[{\cal I}_{n} \otimes {\cal I}_{\bar{n}} \otimes S^{-1}\right]}_{{\cal G}} \otimes f_i \otimes f_j.
\label{eq:fact_outline}
\end{eqnarray}
\begin{itemize}

\item In the first stage of the analysis, full QCD is matched onto an effective field theory which contains fields with the following momentum scalings:
\bea
p_n \sim M (\eta^2, 1, \eta), \qquad p_\bn \sim M (1,\eta^2,\eta), \qquad p_{s}\sim M(\eta,\eta,\eta), \qquad \eta \sim\frac{p_T}{M}, \nn
\eea
corresponding to the n-collinear, $\bn$-collinear, and soft modes respectively. $M$ denotes the mass of either an electroweak gauge boson or the Higgs.  We consider only the leading operators in the $\eta$ expansion, which encode the most singular emissions coming from soft and collinear particles.  The relevant operators for Higgs and electroweak gauge boson production are respectively
\begin{eqnarray}
{\cal O}(\omega_1, \omega_2) &=& g_{\mu \nu}h \> T\{ \text{Tr} \Big [S_n(g B_{n\perp}^\mu)_{\omega_1}S_n^\dagger S_\bn(g B_{\bn \perp}^\nu)_{\omega_2}S_\bn^\dagger \Big ] \}, \nonumber \\
{\cal O}^{Ki}_\mu (\omega_1,\omega_2) &\equiv & (  \bar{\xi}_i W  )_{\bn,\omega_2} T[ S_\bn \Gamma_\mu^{Ki} S_n^\dagger ] ( W^\dagger \xi_i )_{n,\omega_1},
\end{eqnarray}
where the $\Gamma_\mu^{Ki}$ denotes the  Dirac structure, the index $K$ runs over the vector and axial-vector Dirac structures, and the index $i$ runs over the quark flavors.  The $B_{n\perp}^\mu$ and $ B_{\bn \perp}^\nu$ fields  denote collinear-gluon field strengths \cite{Marcantonini:2008qn} dressed with collinear Wilson lines in the $n$ and $\bn$ directions respectively, $\xi_i$ denotes a collinear quark field of flavor $i$, and $W$ denotes a collinear Wilson line.  The $\omega_{1,2}$ are label momenta that give the large light-cone components of the collinear fields.

\item Using the soft-collinear decoupling property of SCET$p_T$, the matrix element in SCET$p_T$ is decoupled into the $n$ and $\bn$ collinear iBFs, $B_n$ and $B_\bn$ respectively, and a soft function as seen in the  fifth line of Eq.~(\ref{fac-schem}). 

\item The $B_n$ and $B_\bn$  iBFs are defined with a zero-bin subtraction \cite{Manohar:2006nz}. Using the equivalence between zero-bin and soft subtractions, explicitly shown at one-loop in our previous work~\cite{Mantry:2009qz} and studied elsewhere in the literature~\cite{Lee:2006nr,Idilbi:2007ff, Idilbi:2007yi}, we can write $B_n B_{\bn}  S \to  \tilde{B}_n \tilde{B}_{\bn}S^{-1}$ where the iBFs $\tilde{B}_{n,\bn}$ are defined without the soft zero-bin subtraction and $S^{-1}$ is the iSF. 

\item In the last step, for $p_T \gg \Lambda_{QCD}$,   the iBFs  are  matched onto standard PDFs  via the  schematic matching equation
\bea
\label{ibf-pdf-1}
\tilde{{B}}_{n,\bn} &=&{\cal I}_{n,\bn} \otimes f .
\eea
The matching coefficients ${\cal I}_{n,\bn}$   are grouped together with the iSF to form a Transverse Momentum Function (TMF) ${\cal G}$ as shown in the last line of Eq.~(\ref{fac-schem}). The TMF encodes the physics of the  soft and collinear emissions from initial-state partons  in the effective theory. For $p_T \sim \Lambda_{QCD}$, the OPE in $\Lambda_{QCD}/p_T$ breaks down so that the perturbative matching in Eq.(\ref{ibf-pdf-1}) is no longer valid. However, in this case one can view Eq.~(\ref{ibf-pdf-1}) simply as an equation that defines  new non-perturbative functions ${\cal I}_{n,\bn}$. Thus,
for $p_T \sim \Lambda_{QCD}$, ${\cal G}$ appears as a new non-pertubative TMF with a well-defined field-theoretic definition and computable anomalous dimension and  can be modeled and extracted from data.

\end{itemize}
The above derivation relies on the cancellation of Glauber mode contributions, as in the standard approach \cite{Collins:1984kg}, to the final observable which measures the $p_T$ of the final-state color-neutral particle. An explicit demonstration of this cancellation using an effective field theory language remains to be shown. For Higgs production, our previous work~\cite{Mantry:2009qz} showed that the differential distribution in transverse momentum and rapidity of the Higgs is given by
\bea
\label{intro-fac}
\frac{d^2\sigma_h}{d\text{p}_T^2 \>dY} &=& \frac{\pi^2}{4(N_c^2-1)^2 Q^2 }     \int_0^1\frac{dx_1}{x_1}\int _0^1\frac{dx_2}{x_2} \int_{x_1}^1\frac{dx'_1}{x'_1}\int _{x_2}^1 \frac{dx'_2}{x'_2}\nn \\
&\times&  H_h(x_1, x_2,\mu_Q;\mu_T){\cal G}_h^{ij}(x_1,x_1',x_2,x_2',p_T,Y,\mu_T)f_{i/P}(x_1' ,\mu_T) f_{j/P}(x_2' ,\mu_T),
\eea
where the we have introduced ${\cal G}_h^{ij}$, theTMF for Higgs boson production, which is given by 
\bea
\label{intro-G}
{\cal G}_h^{ij}(x_1,x_1',x_2,x_2',p_T,Y,\mu_T)&=&  \int dt_n^+ \int  dt_\bn^-\> \int \frac{d^2b_\perp  }{(2\pi)^2}\> J_0 (b_\perp \>p_T)\;g^\perp_{\alpha \sigma} g^\perp_{\beta \omega}\nn \\
&\times&  \> {\cal I}_{n;gi}^{\alpha\beta } (\frac{x_1}{x_1'}, t_n^+,b_\perp,\mu_T)\> {\cal I}_{\bn;gj}^{\sigma \omega} (\frac{x_2}{x_2'}, t_\bn^-,b_\perp,\mu_T) \nn \\
&\times&  {\cal S}_{gg}^{-1}(x_1 Q-e^{Y}\sqrt{\text{p}_T^2+m_h^2}-\frac{t_\bn^-}{x_2 Q}, x_2 Q-e^{-Y}\sqrt{\text{p}_T^2+m_h^2}- \frac{t_n^+}{x_1 Q},b_\perp,\mu_T). \nn \\
\eea
Detailed definitions of the various objects appearing in the above equation are given in sections \ref{giBF} and \ref{iSF}. The integrations over impact parameter $b_{\perp}$ can be explicitly performed to produce an expression written completely in momentum space~\cite{Mantry:2009qz}.  Similarly, the integrations over the residual light-cone momentum components $t_n^+,t_{\bn}^-$ can be performed. For clarity, the explicit expression that results after performing the indicated integrations is shown in Section~\ref{sec:analytic}.

In this paper, we also give a detailed derivation of electroweak gauge boson transverse momentum and rapidity distributions in appendix \ref{appex-1}. The final result is given by 
\bea
\label{intro-DY}
\frac{d^2\sigma}{dp_T^2\> dY}&=& \frac{\pi^2 }{N_c^2}   \int_0^1 dx_1 \int_0^1 dx_2\int_{x_1}^1 \frac{dx_1'}{x_1'} \int_{x_2}^1 \frac{dx_2'}{x_2'}   \nn \\
&\times&   H_Z^{q}(x_1x_2Q^2,\mu_Q;\mu_T) \>{\cal G}^{qrs}(x_1,x_2,x_1',x_2',p_T,Y,\mu_T) f_r(x_1',\mu_T)  f_s(x_2',\mu_T),\nn \\  
\eea
where the TMF function ${\cal G}^{qrs}$
\bea
\label{intro-DY-2}
 &&{\cal G}^{qrs}(x_1,x_2,x_1',x_2',p_T,Y,\mu_T)=  \int \frac{d^2b_\perp}{(2\pi)^2} J_0\big [b_\perp p_T\big ]\>\int dt_n^+ dt_\bn^- \> {\cal I}_{n;q r}(\frac{x_1}{x_1'}, t_n^+,b_\perp,\mu_T)\>{\cal I}_{\bn;\bar{q} s}(\frac{x_2}{x_2'}, t_\bn^-,b_\perp,\mu_T)\nn \\
&\times& {\cal S}^{-1}_{qq}(x_1 Q-e^{Y}\sqrt{\text{p}_T^2+M_Z^2}-\frac{t_\bn^-}{x_2 Q}, x_2 Q-e^{-Y}\sqrt{\text{p}_T^2+M_Z^2}- \frac{t_n^+}{x_1 Q},b_\perp,\mu_T),\nn \\
\eea
and detailed definitions of the various objects can be found in appendix \ref{appex-1}. One can straightforwardly generalize these results to distributions that are differential in the final state leptons.

 We briefly address the appropriate choices for the scales $\mu_Q$ and $\mu_T$ that appear in these equations, as this issue arises later when detailed numerical results are presented.  
These scales should be chosen to minimize logarithms that appear in the expressions for the hard function $H$ and the TMF.  A detailed study of this issue was given in our previous 
work~\cite{Mantry:2009qz}; we summarize the results of this analysis here.  The scale $\mu_Q$ which appears in the hard function should be chosen as $\mu_Q \sim x_1 x_2 Q \sim M_Z$.  The scale $\mu_T$ appearing in the TMF should be chosen to be $\mu_T \sim p_T$.  We note that the TMF is independent of the quantities $t^{+}_n$, $t^{-}_{\bar{n}}$ and $b_\perp$ which are integrated over. The TMF is only sensitive to the transverse momentum and rapidity constraint on the final state Higgs or electroweak gauge boson making $\mu_T \sim p_T$ the natural scale choice.

\section{Fixed-order results}
\label{sec:fixedorder}

In the following sections we derive fixed-order results for the various components of the factorization theorems in Eqs.~(\ref{intro-fac}) and~(\ref{intro-DY}).  We provide mostly outlines of the necessary perturbative calculations, as details of the derivations were already presented in our previous work~\cite{Mantry:2009qz} for the case of Higgs production.

\subsection{Matching Coefficients}
\label{sec:hard}

We begin with the matching coefficients that arise when matching the full QCD current onto operators in SCET$_{p_T}$. The vector-boson current in full QCD is
\bea
J_{Z;\mu} &=& \sum_{i} \big (J^{Vi}_{Z;\mu} + J^{Ai}_{Z;\mu} \big ),
\eea
where $J^{Vi}_{Z;\mu}$ and $J^{Ai}_{Z;\mu}$ are vector and axial-vector currents for the i-th quark flavor
\bea
\label{couplings}
J^{Vi}_{Z;\mu} &=& g_V^i \bar{q}_i \gamma_\mu  q_i, \qquad  J^{Ai}_{Z;\mu} = g_A^i q_i  \gamma_\mu \gamma_5 q_i,
\eea
and $g_V^i$ and $g_A^i$ are the vector and axial vector couplings of the $i$-th quark to the vector boson being studied. We have explicitly used the $Z$ current in these equations, but the results hold identically for $\gamma^*$ and $W$ bosons as well.  These currents are matched onto effective operators in SCET$_{p_T}$ as
\bea
\label{match-Z}
J^{Kj}_{Z;\mu} &=& \int d\omega_1 \int d\omega_2 \> C^{K;ji}(\omega_1,\omega_2,\mu ) {\cal O}^{Ki}_\mu(\omega_1,\omega_2,\mu),
\eea
where the index $K$ takes on the values $K=\{V,A \}$ and the indices $i,j$ run over the quark flavors\footnote{There is also a pure gluon SCET$_{p_T}$ operator that should appear on the RHS of Eq.(\ref{match-Z}). However, the contribution of this operator vanishes~\cite{Stewart:2009yx} for Drell-Yan processes and has thus been left out. }.  $C^{K;ji}$ is the Wilson coefficient of this matching.
The effective operators ${\cal O}_\mu^{Ki} (\omega_1,\omega_2,\mu) $ are given by
\bea
{\cal O}_\mu^{Ki} (\omega_1,\omega_2,\mu)(x) &\equiv &(  \bar{\xi}_i W  )_{\bn,\omega_2} T[ S_\bn^\dagger \Gamma_\mu^{Ki} S_n] ( W^\dagger \xi_i )_{n,\omega_1},
\eea
with the Dirac structures $\Gamma_\mu^{K}$ written as
\bea
\label{dirac}
\Gamma_\mu^{Vi} &=& g_V^i \gamma_\mu^\perp, \qquad \Gamma_\mu^{Ai} = g_A^i \gamma_\mu^\perp \gamma_5.
\eea
$S_{n,\bn}$ denote soft Wilson lines:
\bea
S_{n}(x) = P \>\text{exp}\> \Big [i g \int_{-\infty}^0 n\cdot A_s(x+ sn) \Big ], \qquad  S_{\bn}(x) = P \>\text{exp}\> \Big [i g \int_{-\infty}^0 \bn\cdot A_s(x+ s\bn) \Big ].\nn \\
\eea
Note that in the Dirac structures $\Gamma_\mu^{Kq}$ in Eq.(\ref{dirac}) only the perpendicular components of the index $\mu$ contribute due to the equations of motion of the collinear fields in SCET$_{p_T}$: $\bnslash \xi_n=0, \nslash \xi_\bn=0$.   At tree level, the Wilson coefficient arising from matching the current onto the SCET$_{p_T}$ operator is just unity:
\bea
\label{C-1}
C^{V;ji(0)}(\omega_1, \omega_2,\mu) &=& C^{A;ji(0)}(\omega_1, \omega_2,\mu) = \delta_{ij} .
\eea
The result of the one-loop matching is given by 
\bea 
\label{C-2}
C^{V;ji(1)}(\omega_1, \omega_2,\mu) &=& C^{A;ji(1)}(\omega_1, \omega_2,\mu) \nn \\
& =& \delta_{ij} \frac{\alpha_sC_F}{4\pi} \Bigg [ - \ln ^2 \Big (\frac{\mu^2}{-\omega_1\omega_2 -i\epsilon}\Big ) - 3\ln \Big (\frac{\mu^2}{-\omega_1\omega_2-i\epsilon}\Big ) -8 + \frac{\pi^2}{6}\Bigg ]. \nn \\
\eea
The expression for this Wilson coefficient up to order $\alpha_s^2$ can be found in Refs.~\cite{Becher:2006mr,  Idilbi:2006dg, Becher:2007ty}. The hard Wilson coefficient $H_Z^q(\omega_1,\omega_2,\mu_Q;\mu_T) = H_Z^q(\omega_1 \omega_2, \mu_Q;\mu_T)$ that appears in the factorization theorem in Eq.(\ref{intro-DY}) is given by Eqs.(\ref{hard-1}), (\ref{FKL}), and (\ref{hard-2}).
We refer the reader to \cite{Mantry:2009qz} for a similar analysis for the hard coefficient $H_h(x_1x_2Q^2,\mu_Q;\mu_T)$ that appears in the factorization theorem of Eq.(\ref{intro-fac}) for Higgs production.

\subsection{Quark iBF}

\begin{figure}
\includegraphics[scale=1.25]{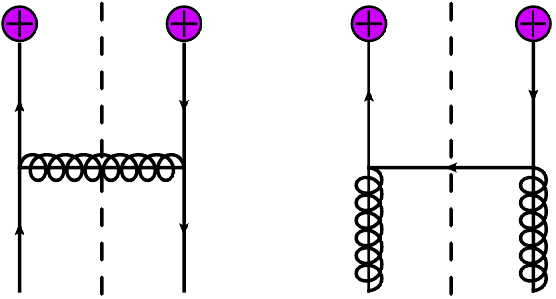}
\caption{Example diagrams for the real emission of a single parton contributing to the next-to-leading order quark iBF.  The purple cross denotes the collinear Wilson lines associated with the $\chi_n$ field.  We note that the momentum $p_1$ is incoming on the left-hand side of the cut and outgoing on the right.}
\label{quarkiBF}
\end{figure}

In this section we give the results for the one-loop calculations of the quark iBFs that appear in the factorization theorem for electroweak gauge boson production. The special case of the quark iBFs with the transverse coordinate $b_\perp=0$ was computed at one loop in ~\cite{Stewart:2010qs}. Our computation closely follows the techniques we established in our previous work~\cite{Mantry:2009qz}.  We compute the $n$-collinear and the $\bn$-collinear quark iBFs by inserting a complete set of states $|X_n\rangle$ and $|X_\bn \rangle $ respectively as
\bea
\tilde{B}^{q}_{n}(x,t,b_\perp,\mu) &=& \frac{1}{2x\,\bn\cdot p_1} \int \frac{db^-}{4\pi} e^{\frac{i t}{2Qx}b^-} \sum_{X_n}  \langle p_1| \bar{\xi}_{nq}W_n(b^-,b_\perp)|X_n\rangle  \langle X_n| \frac{\bnslash}{2}\delta(\bn \cdot {\cal P}- x\>\bn \cdot p_1) W^\dagger_n\xi_{nq}(0) | p_1 \rangle, \nn \\
\tilde{B}^{\bar{q}}_{\bn}(x,t,b_\perp,\mu) &=&  \frac{1}{2x\,n\cdot p_2} \int \frac{db^+}{4\pi} e^{\frac{i t}{2Qx}b^+} \sum_{X_\bn}  \langle p_2|\text{Tr}_{\text{spin}} \Big [  \>\frac{\nslash}{2}(W^\dagger \xi_q)_{\bn}(y)|X_\bn \rangle \langle X_\bn |\delta(-\omega-\bar{{\cal P}}_\bn )(\bar{\xi}_q W)_{\bn}(0) \Big ] | p_2 \rangle, \nn \\
\eea
and then computing the product of matrix elements. The tree level and and virtual corrections are obtained by a perturbative calculation with the choice 
of the vaccum state  $|X_n\rangle =|0\rangle$. The virtual corrections are all scaleless and vanish in pure dimensional regularization. The real emission contributions correspond to choosing final states $|X_n\rangle$ to contain one or more partons.  Example diagrams for the real emission of a single parton are shown in Fig.~\ref{quarkiBF}.  The matching of the iBFs onto the  PDFs is given by
\begin{equation}
\tilde{B}_{n}^q(x, t,b_\perp,\mu) = \int _x^1\> \frac{dz}{z}\> \left\{{\cal I}_{n; qq'} (\frac{x}{z}, t,b_{\perp}, \mu) \>f_{q'}(z,\mu)+
{\cal I}_{n; qg} (\frac{x}{z}, t,b_{\perp}, \mu) \>f_{g}(z,\mu) \right\}
\label{eq:quarkmatch}
\end{equation}
with an analogous equation for the $\bn$-collinear iBF.  The PDFs 
are defined in the usual fashion as
\bea
f_{q}(z,\mu) &=&  \frac{1}{2}\sum_{\text{initial pols.}}  \langle p_1| \bar{\xi}_{nq}W_n(0) \frac{\bnslash}{2}\delta(\bn \cdot {\cal P}- z\>\bn \cdot p_1) W_n^\dagger\xi_{nq}(0) | p_1 \rangle , \nn \\
f_{g}(z,\mu)&=& -\frac{z \,\bn\cdot p_1 }{d-2}\sum_{\text{initial pols.}}\langle p_1 |  \big [  B_{1n \perp}^{A\alpha}  (b^-,b_\perp) \delta(\bar{{\cal P}} -z\bn \cdot p_1 ) B_{1n \perp \alpha}^A  (0)  \big ] | p_1 \rangle .
\eea
In the rest of this section we give results for the $n$-collinear iBF. Analogous results hold for the $\bn$-collinear iBF. At tree level the $n$-collinear iBF is given by
\bea
\tilde{B}^{q(0)}_{n}(x,t,b_\perp,\mu) &=& \delta(t) \delta(1-x).
\eea
At the next order, the emission of a single parton into the final state from the iBF has two contributions:
\bea
\tilde{B}_n^{qR(1)}(x,t,b_\perp,\mu) &=& \tilde{B}_n^{qqR(1)}(x,t,b_\perp,\mu) + \tilde{B}_n^{qgR(1)}(x,t,b_\perp,\mu),
\eea
where $ \tilde{B}_n^{qqR(1)}$ and $ \tilde{B}_n^{qgR(1)}$ correspond  to the first and second diagrams in Fig.~\ref{quarkiBF} respectively.
The contribution of a single gluon emission by an initial state quark to the n-collinear quark iBF is given by the first diagram of Fig.~\ref{quarkiBF}.  All other diagrams do not contribute if a physical polarization sum is used for the final-state gluon. At the level of the integrand, the result for this diagram is
\bea
\tilde{B}_n^{qqR(1)}(x,t,b_\perp,\mu) &=& \frac{g^2 C_F}{2xQ} \int \frac{d^dk}{(2\pi)^{d-1}} \theta(k^0) \delta(k^2)  \delta[Q(1-x)-k^-]\delta[\frac{t}{xQ}-k^+]e^{i\vec{k}\cdot \vec{b}_\perp}\frac{1}{(k -p_1)^4}\nn \\
&\times& \Big [-g_{\mu \nu} + \frac{k_\mu \bn_\nu +k_\nu \bn_\mu}{\bn \cdot k}\Big ]\>\text{Tr}\> \Big [p_1 \!\!\!\!\!\slash \>\gamma^\mu( k\!\!\!\!\slash \>-p_1 \!\!\!\!\!\slash\>)\frac{\bnslash}{2}(p_1 \!\!\!\!\!\slash \>- k\!\!\!\!\slash \>)\gamma^\nu  \Big ],
\eea
where $C_F=4/3$ and we have introduced $Q=\bn \cdot p_1$.  This can be computed in pure dimensional regularization to give
\bea
\tilde{B}_n^{qqR(1)}(x,t,b_\perp,\mu)&=& \frac{\alpha_s C_F}{2\pi} \frac{e^{\epsilon \gamma_E}}{\Gamma(1-\epsilon)} {_{0}F_{1}}\Big (1-\epsilon;-\frac{b_\perp^2 t(1-x)}{4x} \Big) \nn \\
&\times& \frac{1}{\mu^2}\Big [ \frac{\mu^2}{t}\Big ]^{1+\epsilon} \Bigg [(1-\epsilon) x^\epsilon (1-x)^{1-\epsilon} + \frac{2x^{1+\epsilon}}{(1-x)^{1+\epsilon}} \Bigg ].
\eea
We now expand this expression in $\epsilon$ and use the matching condition of Eq.~(\ref{eq:quarkmatch}).  In dimensional regularization, the only contribution to the PDF is at leading-order, as higher-order corrections are scaleless.  Since $f(x) = \delta(1-x)$, we have
\begin{equation}
{\cal I}_{n;qq}(x,t,b_{\perp},\mu) = \left[ \tilde{B}_n^{qq}(x,t,b_{\perp},\mu ) \right]_{\text{finite part in dim. reg.}}.
\end{equation}
We have also performed a calculation where collinear divergences are regulated by introducing an off-shellness for the initial quark, and have obtained identical results for the matching coefficients.  We find the following expression for the matching coefficient:
\begin{eqnarray}
{\cal I}_{n;qq}^{(1)}(x,t,b_{\perp},\mu) &=& \frac{\alpha_s C_F}{2\pi}\Bigg \{\delta(t) \Bigg [-\frac{\pi^2}{6}\delta(1-x) -\frac{1+x^2}{1-x}\ln x +(1-x) \Bigg ] \nn \\
&+&\delta(t) \Big [\frac{\ln (1-x)}{1-x} \Big ]_+(1+x^2) + \frac{2}{\mu^2}\Big [ \frac{\ln (t/\mu^2)}{t/\mu^2}\Big ]_+\delta(1-x)\nn \\
&+&\frac{1}{\mu^2}\Big [ \frac{\mu^2}{t}\Big ]_+\frac{1+x^2}{(1-x)_+} {_{0}F_{1}}\Big (1;-\frac{b_\perp^2 t(1-x)}{4x} \Big)\Bigg \}. \nn \\
\end{eqnarray}

Another contribution to the quark iBF comes from setting the final state $|X_n\rangle $ to a single quark, as shown in the rightmost diagram of Fig.~\ref{quarkiBF}.  This contribution matches to the gluon PDF, and generates the $qg \to Vg$ partonic channel that contributes to electroweak gauge boson production.  The integrand level expression for this diagram is given by
\bea
\tilde{B}_n^{qgR(1)}(x,t,b_\perp,\mu)&=& -\frac{g^2}{2Qx} \int \frac{d^dk}{(2\pi)^{d-1}}\theta(k^0) \delta(k^2) \delta[\frac{t}{xQ}-k^+]\delta[Q(1-x)-k^-]e^{i\vec{k}\cdot \vec{b}_\perp}\frac{1}{(p_1-k)^4}\nn \\
&\times& \Big [ -g_{\mu \nu} + \frac{p_{1\mu}\bn_\nu+p_{1\nu}\bn_{\mu}}{Q} \Big ]\>\text{Tr}\> \Big [k \!\!\!\!\slash \>\gamma^\mu (p_1 \!\!\!\!\!\slash \>- k\!\!\!\!\slash \>)\frac{\bnslash}{2}( k\!\!\!\!\slash \>-p_1 \!\!\!\!\!\slash\>)\gamma^\nu  \Big ].
\eea
In dimensional regularization this becomes
\bea
\tilde{B}_n^{qgR(1)}(x,t,b_\perp,\mu)&=& \frac{\alpha_s}{4\pi} \frac{e^{\epsilon \gamma_E}}{\Gamma(1-\epsilon)} {_{0}F_{1}}(1-\epsilon; -\frac{b_\perp^2 t(1-x)}{4x})\nn \\
&\times& \frac{1}{\mu^2}\Big [\frac{\mu^2}{t}\Big ]^{1+\epsilon}\Big [ 2x^{2+\epsilon}(1-x)^{-\epsilon} - 2x^{1+\epsilon}(1-x)^{-\epsilon} + (1-\epsilon)x^\epsilon (1-x)^{-\epsilon}\Big ]. \nn \\
\eea
Expanding this result in the limit that $\epsilon \to 0$ and keeping only the finite remainder, we derive the matching coefficient
\bea
{\cal I}_{n;qg}^{(1)}(x,t,b_\perp,\mu) &=&  \frac{\alpha_s}{4\pi} \Bigg \{ \delta(t) \Big [\left\{ (1-x)^2+x^2\right\}\ln \frac{1-x}{x} +1 \Big ] \nn \\
&+&  \frac{1}{\mu^2}\Big [ \frac{\mu^2}{t}\Big ]_+\left\{ (1-x)^2+x^2\right\}  {_{0}F_{1}}\Big (1; -\frac{b_\perp^2 t(1-x)}{4x}\Big)\Bigg \}.
\eea

\subsection{Gluon iBF}
\label{giBF}

\begin{figure}
\includegraphics[scale=1.25]{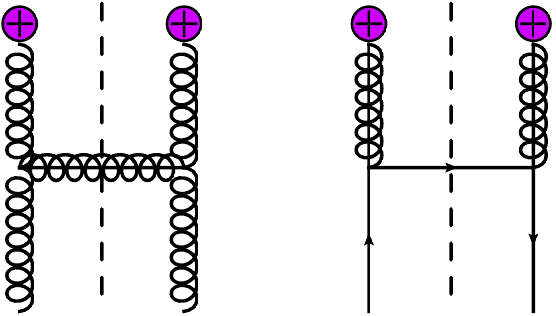}
\caption{Example diagrams contributing to the next-to-leading order gluon iBF.  The purple cross denotes the collinear Wilson lines associated with the $B_{n\perp}$ field.  We note that the momentum $p_1$ is incoming on the left-hand side of the cut and outgoing on the right.}
\label{gluoniBF}
\end{figure}

We now consider the calculation of the gluon iBF.  Although parts of this computation were already presented in Ref.~\cite{Mantry:2009qz}, we include them here for notational consistency and completeness.  Note that in the following we use slightly different definitions for the iBFs compared to that in \cite{Mantry:2009qz}. The n-collinear gluon iBF is given by
\bea
\tilde{B}_{n}^{g;\alpha\beta}(x,t ,b_{\perp},\mu) &=&  -\int \frac{db^-}{4\pi} e^{\frac{i}{2}\frac{t b^-}{x Q}}\sum_{\text{initial pols.}}\sum_{X_n}\langle p_1 |  \big [ g B_{1n \perp \beta}^A  (b^-,b_\perp)|X_n \rangle \nn \\
&\times& \langle X_n |\delta(\bar{{\cal P}} -x_1\bn \cdot p_1 )g B_{1n \perp \alpha}^A  (0)  \big ] | p_1 \rangle ,\nn \\
\eea
where on the right-hand side we use the Fourier transform conjugate variable $t/(x Q)$ compared to $t/Q$ used in \cite{Mantry:2009qz}. The contributing diagrams are shown in Fig.~\ref{gluoniBF}.  In dimensional regularization and with a physical polarization sum used for the gluons, these are the only 
contributions.  The matching equation for the gluon iBF onto the PDFs is given by
\begin{equation}
\tilde{B}_{n;\alpha\beta}^g(x, t,b_\perp,\mu) = \int _x^1\> \frac{dz}{z}\> \left\{{\cal I}^{gg}_{n; \alpha\beta} (\frac{x}{z}, t,b_{\perp}, \mu) \>f_{g}(z)+
{\cal I}^{gq}_{n; \alpha\beta} (\frac{x}{z}, t,b_{\perp}, \mu) \>f_{q}(z) \right\}.
\label{eq:gluonmatch}
\end{equation}

The tree-level expression for the iBF is given by
\bea
\tilde{B}_{n}^{(0)\alpha\beta}(x,t ,b_{\perp},\mu) &=& g^2 g_\perp^{\alpha \beta} \delta(t)\delta(1-x).
\eea
The virtual corrections to the iBF are scaleless and vanish in pure dimensional regularization. At the next order beyond tree-level, the real emission of a single parton from the iBF has two types of contributions:
\bea
\tilde{B}_{n;\alpha\beta}^{gR(1)}(x,t, b_{\perp},\mu) &=& \tilde{B}_{n;\alpha\beta}^{ggR(1)}(x,t, b_{\perp},\mu)  + \tilde{B}_{n;\alpha\beta}^{gqR(1)}(x,t, b_{\perp},\mu), 
\eea
where $\tilde{B}_{n;\alpha\beta}^{ggR(1)}$ and $\tilde{B}_{n;\alpha\beta}^{gqR(1)}$ correspond to the emission of a real gluon and a real quark in the final state respectively. We begin by considering the contribution arising from single-gluon emission into the final state, which corresponds to the leftmost diagram in Fig.~\ref{gluoniBF}.  The result can be expanded in terms of two form factors,
\begin{equation}
\tilde{B}_{n;\alpha\beta}^{gg}(x,t, b_{\perp},\mu) = \mathcal{F}_1^{g;\epsilon} (x,t,b_{\perp},\mu) g_{\perp}^{\alpha\beta} +\mathcal{F}_2^{g;\epsilon} (x,t,b_{\perp},\mu) \left[ g_{\perp}^{\alpha\beta} +(d-2)\frac{\vec{b}_{\perp}^{\alpha} \vec{b}_{\perp}^{\beta}}{b_{\perp}^2}\right], \nn \\
\label{ffdecomp}
\end{equation}
where $\mathcal{F}_{1,2} $ are given by
\begin{eqnarray}
\label{f1f2}
\mathcal{F}_1^{gg;\epsilon} &=& g^2 \frac{\alpha_s C_A}{\pi} \frac{e^{\epsilon \gamma_E}}{\Gamma(1-\epsilon)}\frac{1}{\mu^2} \left[ \frac{\mu^2}{t}\right]^{1+\epsilon} x^{1+\epsilon}(1-x)^{-\epsilon} {_{0}F_{1}}\Big (1-\epsilon; -\frac{b_\perp^2 t(1-x)}{4x}\Big)  \nonumber \\ 
&\times& \left\{1-x+\frac{1-x}{x^2}+\frac{1}{1-x}\right\}, \nonumber \\
\mathcal{F}_2^{gg;\epsilon} &=& g^2 \frac{\alpha_s C_A}{4\pi} \frac{e^{\epsilon \gamma_E}}{\Gamma(3-\epsilon)} \left[ \frac{\mu^2}{t}\right]^{\epsilon}
\frac{(1-x)^{2-\epsilon}}{x^2} {_{0}F_{1}}\Big (3-\epsilon; -\frac{b_\perp^2 t(1-x)}{4x}\Big).
\end{eqnarray}
The quantity ${\cal F}_2$ is finite, and we can immediately set $\epsilon$ to zero.  ${\cal F}_1$ must be expanded in distributions.  We do so and drop the pole terms, as explained in the section on the quark iBF.  For the gluon iBF we utilize the matching equation
\bea
\tilde{B}_{n}^{g;\alpha\beta}(x,t ,b_{\perp},\mu) &=& \int_x^1\frac{dz}{z} {\cal I}^{\alpha\beta}_{ng;i}(\frac{x}{z},t,b_\perp,\mu ) f_i(z,\mu),
\label{eq:gluonmatch}
\eea
which leads to the result
\begin{equation}
{\cal I}_{n;gg}^{\alpha\beta}(x,t, b_{\perp},\mu) = \mathcal{F}_1^{gg} (x,t,b_{\perp},\mu) g_{\perp}^{\alpha\beta} +\mathcal{F}_2^{gg} (x,t,b_{\perp},\mu) \left[ g_{\perp}^{\alpha\beta} +(d-2)\frac{\vec{b}_{\perp}^{\alpha} \vec{b}_{\perp}^{\beta}}{b_{\perp}^2}\right],
\end{equation}
with
\begin{eqnarray}
\mathcal{F}_{1}^{gg} &=& g^2 C_A \frac{\alpha_s}{\pi} \left\{ -\frac{\pi^2}{12} \delta(t) \, \delta(1-x) + \delta(1-x) \frac{1}{\mu^2}\left[\frac{\mu^2}{t}\text{ln}	\left(\frac{t}{\mu^2}\right)\right]_+  \right. \nonumber \\
	&+&\delta(t)\>x\left[ \left( 1-x +\frac{1-x}{x^2}\right) \text{ln}(1-x) + \left[ \frac{\text{ln}(1-x)}{1-x}\right]_+ \right] \nonumber \\
         &+& \left. \frac{1}{\mu^2}\left[ \frac{\mu^2}{t}\right]_+ x \left[1-x +\frac{1-x}{x^2}+\frac{1}{[1-x ]_+}\right] {_{0}F_{1}} \left(1; -\frac{b_{\perp}^2 t (1-x )}{4x}\right) \right. \nn \\
	&-&\left.   \delta(t)\> x\>\text{ln} x  \left[1-x +\frac{1-x}{x^2}+\frac{1}{[1-x ]_+}\right]  \right\}, \nonumber \\
\mathcal{F}_2^{gg} &=& g^2 C_A \frac{\alpha_s}{8\pi} \frac{(1-x )^2}{x^2} b_\perp^2 \>{_{0}F_{1}} \left(3 ; -\frac{b_{\perp}^2 t (1-x )}{4x}\right).
\end{eqnarray}

There is another contribution to the gluon iBF coming from radiating a quark into the final state, which comes from the rightmost diagram of Fig.~\ref{gluoniBF}.  This matches onto the quark PDF, and generates the matching coefficient ${\cal I}_{n;gq}$.  As with $\tilde{B}_{n;\alpha \beta}^{gg}$, this can be expanded in two form factors, leading to the generic form
\bea
\tilde{B}_{n;\alpha \beta}^{gq} (x,t,b_\perp,\mu) &=& {\cal F}_1^{gq} (x,t,b_\perp,\mu) g_{\alpha \beta}^{\perp} + {\cal F}_2^{gq}(x,t,b_\perp,\mu) \big [ g_{\alpha \beta}^\perp + (d-2) \frac{\vec{b}^\perp_\alpha \vec{b}^\perp_\beta}{\vec{b}^2}\big ].
\eea
After performing the integrations and expanding in $\epsilon$ as in the previous contributions, and utilizing the matching in Eq.~(\ref{eq:gluonmatch}), we 
find the following expressions for the two form factors:
\bea
{\cal F}_{1}^{gq} &=& g^2 \frac{\alpha_s C_F}{2\pi}\left\{\frac{1}{\mu^2}\Big [ \frac{\mu^2}{t}\Big ]_+ \frac{(1-x)^2 +1}{x} {_{0}F_{1}}(1; -\frac{b_\perp^2 t(1-x)}{4x}) \right. \nn \\
 &-&  \left.\delta(t) \left[\frac{(1-x)^2 +1}{x} \ln \frac{ 1-x}{ x}   -2 \frac{1-x}{x}   \right]\right\}, \nn \\
{\cal F}_2^{gq} (x,t,b_\perp,\mu) &=& g^2 \frac{\alpha_s C_F}{2\pi} \frac{(1-x)^2b_\perp^2}{4x^2}  {_{0}F_1}(3; -\frac{b_\perp^2 t(1-x)}{4x}). 
\eea

\subsection{Inverse Soft Functions}
\label{iSF}

We now discuss the computation of the inverse soft functions that appear in the factorization theorem for both Higgs and electroweak gauge boson production.  They serve to remove an overcounting of soft emissions that occur when the iBFs are inserted into the factorization theorem, and are needed to correctly match the fixed-order cross section.  The relevant iSF for gauge boson production is
\bea
 {\cal S}^{-1}_{qq}(\tilde{\omega}_1,\tilde{\omega}_2,b_\perp,\mu) &=& \int \frac{db^+db^-}{16\pi^2} e^{\frac{i}{2}\tilde{\omega}_1b^+} e^{\frac{i}{2}\tilde{\omega}_2b^-} S_{qq}^{-1}(b^+,b^-,b_\perp,\mu),
\eea
where $S_{qq}(b,\mu)$ is a vacuum matrix element of soft Wilson lines given by
\bea
S_{qq}(b,\mu) &=&\sum_{X_s} \text{Tr} \langle 0 |\bar{T} [ S_n^\dagger S_\bn ] (z)|X_s\rangle \langle X_s| \> T [ S_\bn^\dagger S_n ](0) |0\rangle .
\eea
We have inserted a complete set of final states  $| X_s \rangle$.  The relevant formulae for the iSF in Higgs production are 
\bea
\mathcal{S}^{-1}_{gg}(\tilde{\omega}_1,\tilde{\omega}_2,b_{\perp},\mu) &=& \int \frac{db^+ db^-}{16\pi^2} e^{ib^+ \tilde{\omega}_1 /2} e^{i b^- \tilde{\omega}_2 /2} S_{gg}^{-1}(b^+ ,b^-, b_{\perp}), \nn \\
S_{gg}(b,\mu) &=& \sum_{X_{s}} \langle 0 | \bar{T} \left[\text{Tr} \left ( S_\bn T^D S_\bn^\dagger S_n T^C S_n^\dagger \right ) (b)\right] |X_{s}\rangle \langle X_{s} | T \left[\text{Tr} \left ( S_n T^C S_n^\dagger S_\bn T^D S_\bn^\dagger \right ) (0)\right] | 0 \rangle . \nn \\
\eea
The tree level result for the quark iSF is given by
\bea
 {\cal S}_{qq}^{-1(0)}(\frac{t_{\bn}^{max}-t_\bn^-}{x_2Q}, \frac{t_n^{max}-t_n^+}{x_1Q},b_{\perp},\mu ) &=& N_c \>x_1 x_2 \>Q^2 \delta (t_\bn^{max}- t_\bn^-)\delta (t_n^{max}- t_n^+), \nn \\
 \eea
where we have used the same arguments in the iSF that appear in the factorization theorem so that
\bea
\label{tntbnmax}
t_\bn^{max} &\equiv& x_1 x_2 Q^2 - x_2(M^2-u), \qquad t_\bn^{max} \equiv x_1 x_2 Q^2 - x_1(M^2-t).
\eea
$u$ and $t$ are the hadronic Mandelstam invariants that appear in the factorization theorem derived in appendix \ref{appex-1}.  The result for the gluon iSF at tree level can be obtained via a simple scaling of the quark result:
\begin{equation}
{\cal S}_{gg}^{-1(0)} = \frac{N_C^2-1}{4 N_C} {\cal S}_{qq}^{-1(0)}.
\end{equation}

\begin{figure}
\includegraphics[scale=1.25]{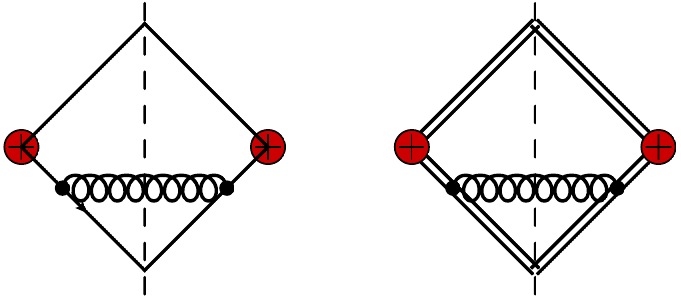}
\caption{Example diagrams contributing to the next-to-leading order iSFs for both electroweak gauge boson production (left) and Higgs production (right).  The lines at each vertex schematically denote the soft Wilson lines associated appearing in the definition of the iSFs.   }
\label{SCETsoft}
\end{figure}

As with the iBFs, the contributions from the higher-order virtual corrections with  $| X_s \rangle = |0 \rangle$ are scaleless, and vanish in dimensional regularization.  The only contributions come from real-emission diagrams.  Examples are shown in Fig.~\ref{SCETsoft} for both vector boson and Higgs production.  For gauge boson production, the contribution of real gluon emission to the iSF takes the integrand level form
\bea
\mathcal{S}_{qq}^{-1 R(1)}(\frac{t_{\bn}^{max}-t_\bn^-}{x_2Q}, \frac{t_n^{max}-t_n^+}{x_1Q},b_{\perp},\mu ) &=& 4g^2 N_c C_F \int \frac{d^dk}{(2\pi)^{d-1}}\frac{\delta (k^2) \theta(k^0) e^{i\vec{k}_\perp \cdot \vec{b}_\perp}}{k^+k^-}\nn \\
&\times& \delta\Big [\frac{t_{\bn}^{max}-t_\bn^-}{x_2Q}-k^-\Big ]\delta\Big [\frac{t_{n}^{max}-t_n^+}{x_1Q}-k^+\Big ].
\eea
Evaluating this in dimensional regularization and dropping the pole terms yields
\begin{eqnarray}
\label{iSFreal-2}
S_{qq}^{-1R (1)} &=& -N_c \frac{\alpha_s C_F}{\pi} \hat{Q}^2 \left\{ -\frac{\pi^2}{12} \delta(t_{\bar{n}}^{max}-t_{\bar{n}}^-)  \delta(t_n^{max}-t_n^+) +\frac{1}{2} 	\delta(t_{\bar{n}}^{max}-t_{\bar{n}}^-)\delta(t_n^{max}-t_n^+) \text{ln}^2 \frac{\hat{Q}^2}{\mu^2} \right. \nonumber \\
	&+& \delta(t_{\bar{n}}^{max}-t_{\bar{n}}^-) \frac{1}{\hat{Q}^2}\left[ \frac{\hat{Q}^2}{t_n^{max}-t_n^+}\right]_+ \text{ln} \frac{\hat{Q}^2}{\mu^2} + \delta(t_n^{max}-t_n^+)  \frac{1}{\hat{Q}^2}\left[ \frac{\hat{Q}^2}{t_{\bar{n}}^{max}-	t_{\bar{n}}^-} \right]_+ \text{ln} \frac{\hat{Q}^2}{\mu^2} \nonumber \\
	&+& \delta(t_{\bar{n}}^{max}-t_{\bar{n}}^-) \frac{1}{\hat{Q}^2}\left[ \frac{\hat{Q}^2}{t_n^{max}-t_n^+}\text{ln}\frac{t_n^{max}-t_n^+}{\hat{Q}^2}\right]_+ + \delta(t_n^{max}-t_n^+)  \frac{1}{\hat{Q}^2}\left[ \frac{\hat{Q}^2}{t_{\bar{n}}^{max}-	t_{\bar{n}}^-}\text{ln}\frac{t_{\bar{n}}^{max}-t_{\bar{n}}^-}{\hat{Q}^2}\right]_+ \nonumber \\
	&+&  \left. \frac{1}{\hat{Q}^4}\left[ \frac{\hat{Q}^2}{t_n^{max}-t_n^+}\right]_+  \left[ \frac{\hat{Q}^2}{t_{\bar{n}}^{max}-t_{\bar{n}}^-} \right]_+  {_{0}F_{1}}\left( 1; -\frac{b_{\perp}^2 (t_n^{max}-t_n^+)	(t_{\bar{n}}^{max}-t_{\bar{n}}^-)}{4\hat{Q}^2}\right) \right\}.
\label{softfunc}
\end{eqnarray}
We have used $\hat{Q}^2 = x_1 x_2 Q^2$ to simplify the notation.  The result for the iSF in Higgs production at this order can be obtained via the scaling
\begin{equation}
{\cal S}_{gg}^{-1R(1)} = \frac{(N_C^2-1)C_A}{4 N_C\,C_F} {\cal S}_{qq}^{-1R(1)}.
\end{equation}

\section{Running}
\label{sec:running}

In the section we briefly summarize the structure of the running of the various objects that appear in the factorization theorems
for the transverse momentum and rapidity distributions. For a  more detailed discussion of the running structure see Ref.~\cite{Mantry:2009qz} where
the case of Higgs production was studied in great detail. The overall structure of factorization is similar for both Drell-Yan and 
Higgs production and is schematically characterized by a hard function $H$, a transverse momentum  function (TMF)
${\cal G}$, and the standard initial state PDFs. The TMF ${\cal G}$
is a convolution over two iBFs and an iSF. It is only the specific forms of the hard and transverse momentum functions and the 
the type of parton PDFs that dominate that differ in Drell-Yan and the Higgs production processes. In this section we summarize the
running structure for the case of the $Z$-boson distribution function.

The evolution equations for $H^q_Z(\hat{Q}^2,\mu)$ are diagonal in flavor (for a more detailed discussion see~\cite{Stewart:2009yx, Stewart:2010qs,Stewart:2010pd}), and one can write
\bea
\label{anom-gen-4}
\mu \frac{d}{d\mu} H_Z^q =\, \gamma_{H^q_Z}\>H_Z^q,
\eea
where the anomalous dimension $\gamma_{H^q_Z}$ has the form
\bea
\label{anom-gen-5}
\gamma_{H_{Z}^{q}}  = \Gamma_H^q[\alpha_s] \ln \frac{\hat{Q}^2}{\mu^2} + \gamma_H^q[\alpha_s].
\eea
The first term proportional to $ \ln \frac{\hat{Q}^2}{\mu^2}$ is known as the cusp anomalous dimension and the second term
is the non-cusp piece. The quantities $\Gamma_H[\alpha_s]$ and $\gamma_H[\alpha_s]$ have perturbative expansions in $\alpha_s$  
of the form
\bea
\label{anom-gen-6}
\Gamma_{H}^q [\alpha_s] = \frac{\alpha_s}{4\pi} \Gamma_0^{Hq}+ \Big [  \frac{\alpha_s}{4\pi} \Big ]^2 \Gamma_1^{Hq} +\cdots, \qquad \gamma_{H}^q [\alpha_s] = \frac{\alpha_s}{4\pi} \gamma_0^{Hq} + \Big [  \frac{\alpha_s}{4\pi} \Big ]^2 \gamma_1^{Hq} +\cdots
\eea
The matching from QCD onto $\text{SCET}_{p_T}$ is the same as for the study of threshold resummation in SCET, and the results for the anomalous dimensions can be obtained from previous studies~\cite{Becher:2006mr,Ahrens:2008nc}.  For resummation to NLL accuracy, the following coefficients 
of the expansion are needed:
\bea
\label{anomcoeffs}
\Gamma_0^{Hq} &=& 8 C_F, \qquad \gamma_0^{Hq} = -12 C_F, \nonumber \\
\Gamma_1^{Hq} &= &8 C_F \left[ \left( \frac{67}{9} -\frac{\pi^2}{3}\right) C_A -\frac{10}{9} N_F \right].
\eea
The solution to the evolution of the hard function between the scales $\mu_Q\sim \hat{Q}$ and $\mu_T\sim p_T$ and takes the form
\bea
H^q_Z(\hat{Q}^2, \mu_Q;\mu_T) &=& U_{H^q_Z}(\hat{Q}^2,\mu_Q,\mu_T)H^q_Z(\hat{Q}^2, \mu_Q) ,
\eea
where $U_{H^q_Z}(\hat{Q}^2,\mu_Q,\mu_T)$ denotes the evolution factor and sums the logarithms of $\hat{Q}^2/p_T^2$.

The TMF function ${\cal G}^{qrs}$ that appears in the factorization theorem as seen in Eq.(\ref{intro-DY-2}) is evaluated at the scale $\mu_T \sim p_T$. The PDFs are evolved via standard DGLAP equations up to the scale $\mu_T \sim p_T$, summing up the remaining logarithms of $\Lambda_{QCD}/p_T$. For the running between $\mu_Q\sim \hat{Q}\sim M$ and $\mu_T \sim p_T$, one can also consider running of the iBFs and the iSF individually above the $\mu_T \sim p_T$ scale. It was shown in \cite{Mantry:2009qz} that the combined convolution running of the iBFs and the iSF cancels the running of the hard function as required by the scale invariance of the cross-section.  The running of the iBF can be written as
\bea 
\mu \frac{d}{d\mu}B^q(x,t,b_\perp,\mu) &=& \int dt' \gamma_q^B (t-t',\mu) B^q(x,t',b_\perp,\mu),
\eea
where at one loop, the anomalous dimension is given by
\bea
\gamma_q^{B(1)}(t,\mu) &=&  \frac{\alpha_s(\mu)C_F}{\pi} \Big [-\frac{2}{\mu^2}\Big [ \frac{\mu^2}{t}\Big ]_+ + \frac{3}{2}\delta(t) \>\Big ],
\eea
which is the same as what was found for the iBF with $b_\perp=0$ in \cite{Stewart:2010qs}.
The anomalous dimension of the iSF ${\cal S}^{-1}_{qq}$ is determined by the equation
\bea 
\mu \frac{d}{d\mu}{\cal S}^{-1}_{qq}(\frac{t_\bn}{Q},\frac{t_n}{Q},b_\perp,\mu) &=&\frac{1}{Q^2} \int dt_n'\int dt_\bn' \gamma_{{\cal S}^{-1}_{qq}} (\frac{t_\bn}{Q}-\frac{t_\bn'}{Q},\frac{t_n}{Q}-\frac{t_n'}{Q},\mu){\cal S}^{-1}_{qq}(\frac{t_\bn'}{Q},\frac{t_n'}{Q},b_\perp,\mu), \nn \\
\eea
and at one loop is given by
\bea
\gamma_{{\cal S}^{-1}_{qq}}^{(1)}(\frac{t_\bn}{Q},\frac{t_n}{Q},\mu) &=&  \frac{2 \alpha_s C_F}{\pi} Q^2 \Bigg [ \delta(t_\bn)\frac{1}{Q^2}\Big [\frac{Q^2}{t_n}\Big ]_+  +  \delta(t_n)\frac{1}{Q^2}\Big [\frac{Q^2}{t_\bn}\Big ]_+  + \delta(t_n)\delta(t_\bn) \ln \frac{Q^2}{\mu^2}\Bigg ]. \nn \\
\eea


\section{Analytic expressions for resummed cross sections}
\label{sec:analytic}

Using the fixed-order expressions and renormalization-group evolution of the hard function derived in the previous sections, we can derive the explicit expressions for the differential cross sections which resum large logarithms for low transverse momenta.  We begin by considering the case of Higgs boson production.  The relevant factorization formula is shown in Eq.~(\ref{intro-fac}).  The running of the hard function $H_h$ was described in Section~\ref{sec:running}.  For the transverse momentum function defined in Eq.~(\ref{intro-G}), we must plug in the perturbative expansions for the relevant iBFs and iSF.  These were derived in Section~\ref{sec:fixedorder}.  We note that if tree-level expressions are used for both beam functions and the soft function, then the phase-space constraints force $p_T=0$.  Therefore, the NLO expressions for one of these functions must be used.

We derive here the ${\cal O}(\alpha_s)$ expressions for the differential cross sections, which correspond to the leading-order result for the $p_T$ spectrum.  Upon plugging in the expressions for the iBFs and the iSF, the integrals over $b_{\perp}$, $t_n^+$, $t_{\bn}^-$, $x_1$, and $x_2$ can be performed.  The differential cross section  can be written in the general form
\begin{equation}
\label{hadcross}
\frac{d^2 \sigma_{h,Z}}{dp_T^2 dY} = \sum_{i,j} \int d x_1 d x_2 \, f_{i/P}(x_1) f_{j/P}(x_2) \,\hat{Q}^2 \frac{d^2 \sigma_{h,Z}^{ij}}{d \hat{u} \,d\hat{t}},
\end{equation}
where the subscripts $h,Z$ refer to the Higgs and Z boson production respectively.
We have introduced the usual partonic Mandelstam invariants $\hat{Q}$, $\hat{t}$, and $\hat{u}$.  There are three relevant partonic channels that contribute to Higgs production: $gg \to gh$, $qg \to qh$, and $q\bar{q} \to gh$.  The partonic differential cross section for the $gg$ initial state can be written 
as follows:
\begin{eqnarray}
\frac{d^2 \sigma_h^{gg}}{d \hat{u} \,d\hat{t}} &=& \frac{\pi^2}{192 v^2} \left( \frac{\alpha_s}{\pi}\right)^3 \left\{ -\frac{\hat{Q}^2}{\hat{t}\hat{u}} U_h (\hat{Q}^2,\mu_Q,\mu_T) + \frac{\hat{Q}^2}{\hat{t}\hat{u}} \left[1+\frac{\hat{t}}{\hat{Q}^2} + \frac{\hat{t}^2}{\hat{Q}^4} \right]^2 U_h (m_h^2-\hat{u},\mu_Q,\mu_T) \right. \nn \\
&+&\left.  \frac{\hat{Q}^2}{\hat{t}\hat{u}} \left[1+\frac{\hat{u}}{\hat{Q}^2} + \frac{\hat{u}^2}{\hat{Q}^4} \right]^2 U_h (m_h^2-\hat{t},\mu_Q,\mu_T)
+\frac{2\hat{t}^2+2\hat{u}^2+3\hat{t}\hat{u}+6\hat{Q}^2m_h^2}{\hat{Q}^6} \right\}\nn \\
&\times&  \delta(\hat{Q}^2+\hat{t}+\hat{u}-m_h^2).
\end{eqnarray}
The four terms appearing in this result have a clear origin.  The first term arises from $\mathcal{S}^{-1}_{gg}$, while the second and third come from the NLO expressions for the iBFs.  The last term results from matching the expressions to the fixed-order QCD result.  In the limit that the $p_T$ of the Higgs becomes large, and therefore that $\mu_T$ approaches $\mu_Q$, the evolution factors $U_h \to 1$, and the regular fixed-order QCD result is obtained.  Of the remaining channels, the $qg$ initial state has only a contribution from the NLO iBF, while the $q\bar{q}$ has no contribution from soft or collinear emissions at this order and comes entirely from matching to the fixed-order QCD result.  The explicit expressions are
\begin{eqnarray}
\frac{d^2 \sigma_h^{qg}}{d \hat{u} \,d\hat{t}} &=& -\frac{\pi^2}{864 v^2} \left( \frac{\alpha_s}{\pi}\right)^3 \frac{1}{\hat{u}}\left\{1+\frac{\hat{t}^2}{\hat{Q}^4} \right\} U_h (m_h^2-\hat{u},\mu_Q,\mu_T) \delta(\hat{Q}^2+\hat{t}+\hat{u}-m_h^2), \nn \\
\frac{d^2 \sigma_h^{q\bar{q}}}{d \hat{u} \,d\hat{t}} &=& \frac{\pi^2}{324 v^2} \left( \frac{\alpha_s}{\pi}\right)^3 \frac{1}{\hat{Q}^2}\left\{\frac{\hat{u}^2}{\hat{Q}^4} +\frac{\hat{t}^2}{\hat{Q}^4} \right\} \delta(\hat{Q}^2+\hat{t}+\hat{u}-m_h^2). 
\end{eqnarray}

For vector boson production, we for simplicity explicitly write the result only for on-shell $Z$-boson production with the leptonic phase space integrated over.  Results for $W$ and $\gamma^{*}$ production can be obtained through simple modifications of this formula, as can the results when the leptons are treated differentially.  In this case, two partonic channels contribute, $q\bar{q}$ and $qg$.  The result for the cross section can be written exactly as for the Higgs in Eq.~(\ref{hadcross}).  The partonic channels can be written as follows:
\begin{eqnarray}
\label{eq:anform}
\frac{d^2 \sigma_Z^{q\bar{q}}}{d \hat{u} \,d\hat{t}} &=& \frac{2\pi}{9} \alpha_S \frac{\left[(g_V^{q})^2+(g_A^{q})^2\right]}{4\pi} \text{Br}\left( Z \to l^+l^-\right)
\left\{ -\frac{2}{\hat{t}\hat{u}}U_{H^q_Z} (\hat{Q}^2,\mu_Q,\mu_T) \right. \nn \\
&+&\left. \frac{1}{\hat{Q}^4}\left[ \frac{\hat{t}}{\hat{u}} + 2\frac{\hat{Q}^2}{\hat{u}}+2\frac{\hat{Q}^4}{\hat{t}\hat{u}}\right]U_{H^q_Z} (M_Z^2-\hat{u},\mu_Q,\mu_T)
+\frac{1}{\hat{Q}^4}\left[ \frac{\hat{u}}{\hat{t}} + 2\frac{\hat{Q}^2}{\hat{t}}+2\frac{\hat{Q}^4}{\hat{t}\hat{u}}\right]U_{H^q_Z} (M_Z^2-\hat{t},\mu_Q,\mu_T)\right\}\nn \\
&\times&  \delta(\hat{Q}^2+\hat{t}+\hat{u}-M_Z^2),
\nn \\
\frac{d^2 \sigma_Z^{qg}}{d \hat{u} \,d\hat{t}} &=&  -\frac{\pi}{12} \alpha_S  \frac{\left[(g_V^{q})^2+(g_A^{q})^2\right]}{4\pi} \text{Br}\left( Z \to l^+l^-\right)
\left\{ \frac{1}{\hat{Q}^2\hat{u}}\left[1+2 \frac{\hat{t}}{\hat{Q}^2}+2\frac{\hat{t}^2}{\hat{Q}^4}\right] U_{H^q_Z}(M_Z^2-\hat{u},\mu_Q,\mu_T) \right. \nn \\
&+& \left. \frac{1}{\hat{Q}^4}\left[ \frac{\hat{t}}{\hat{Q}^2}+2\frac{\hat{u}}{\hat{Q}^2}\right] \right\} \delta(\hat{Q}^2+\hat{t}+\hat{u}-M_Z^2),
\end{eqnarray}
with the vector and axial couplings as defined in Eq.~(\ref{couplings}). The remaining partonic channel $\sigma_Z^{gq}$ can be obtained from $\sigma_Z^{qg}$
by the interchange of the $\hat{u}$ and $\hat{t}$ variables.   At the order to which we are working, the explicit form of the evolution factor $U_{H^q_Z}(\xi,\mu_Q,\mu_T)$ appearing in Eq.~(\ref{eq:anform}) is given by:
\begin{eqnarray}
\label{eq:evolution}
U_{H^q_Z}(\xi,\mu_Q,\mu_T) &=& |{\rm exp}\left\{ 2S(\mu_Q,\mu_T) - a_{\Gamma}(\mu_Q,\mu_T) {\rm ln}(-\xi/\mu_Q^2) -a_{\gamma}(\mu_Q,\mu_T)\right\}|^2, \nonumber \\
S(\mu_Q,\mu_T) &=& -\frac{\Gamma_0^{Cq}}{16 \beta_0^2}\left\{ \frac{\left( 1-r+r \, {\rm ln} \, r \right)}{r \, \alpha_S(\mu_Q)} +\frac{\beta_1}{2 \beta_0} {\rm ln}^2 r
	+\left( \frac{\Gamma_1^{Cq}}{4 \Gamma_0^{Cq}}-\frac{\beta_1}{\beta_0}\right)\left( 1-r+{\rm ln}\,r \right) \right\}, \nonumber \\
a_{\Gamma}(\mu_Q,\mu_T) &=& \frac{\Gamma_0^{Cq}}{8\beta_0} {\rm ln} \, r,\;\;\; a_{\gamma}(\mu_Q,\mu_T) = \frac{\gamma_0^{Cq}}{8\beta_0} {\rm ln} \, r,
 \end{eqnarray}
 where $\Gamma_{0,1}^{Cq}\equiv \Gamma_{0,1}^{Hq}/2, \gamma_0^{Cq}\equiv\gamma_0^{Hq}/2$, and $r = \alpha_S(\mu_T)/\alpha_S(\mu_Q)$ and $\beta_{0,1}$ are the coefficients of the QCD beta function in the normalization where $\beta_0 = (11-2 N_F/3)/4$.  The expressions 
 for $\Gamma_{0,1}^{Hq}$ and $\gamma_0^{Hq}$ were given in Eq.~(\ref{anomcoeffs}).  This form of $U_{Q^q_Z}$ can be obtained from the results in Ref.~\cite{Becher:2006mr}.


\section{Comparison with the Collins-Soper-Sterman approach}
\label{sec:css}

We now compare with the standard CSS approach to transverse momentum resummation, and demonstrate that the logarithms resumed are equivalent through next-to-leading logarithmic accuracy (NLL), {\it i.e.}, through next-to-leading order in the resummed exponent.  We outline what further calculations are needed to extend this result to higher orders. 

We begin by comparing the exponents that implement the resummation of large logarithms of the scales $M$ and $p_T$.  The CSS approach writes the transverse momentum distribution as 
\begin{eqnarray}
\frac{d^2 \sigma}{dp_T \,dY} &=& \sigma_0 \int \frac{d^2 b_{\perp}}{(2\pi)^2} e^{-i \vec{p}_T \cdot \vec{b}_{\perp}} \sum_{a,b} \left[ C_a \otimes f_{a/P} \right] (x_A, b_0/b_{\perp} )\left[ C_b \otimes f_{b/P} \right] (x_B, b_0/b_{\perp} ) \nn \\
	&\times& \text{exp}\left\{ \int_{b_0^2/b_{\perp}^2}^{\hat{Q}^2}  \frac{d \mu^2}{\mu^2} \left[ \text{ln}\frac{\hat{Q}^2}{\mu^2} A(\alpha_s (\mu)) +B(\alpha_s (\mu))\right] \right\},
\end{eqnarray}
where we have neglected the remainder term $Y$ which also appears for the purpose of this discussion.  The coefficients $A$, $B$, and $C_a$ have perturbative expansions in the strong coupling constant:
\begin{equation}
\label{csscoeffs}
A = \sum_{n=1} \left(\frac{\alpha_s(\mu)}{\pi}\right) A^{(n)},\;\; B = \sum_{n=1} \left(\frac{\alpha_s(\mu)}{\pi}\right) B^{(n)}, \;\; C_a = \sum_{n=0} \left(\frac{\alpha_s(\mu)}{\pi}\right) C_a^{(n)}.
\end{equation}
The results for these coefficients are well-known in the literature~\cite{Collins:1984kg}.

The exponentiation of low $p_T$ logarithms in our approach is accomplished via the RG evolution of the hard function $H$ from $\mu_Q \sim M$ to 
$\mu_T \sim p_T$.  As noted before in Section~\ref{sec:hard}, the matching coefficient here is the same as that for inclusive production of vector bosons or the Higgs.  Therefore, we can take the solution for the hard-function evolution from the literature~\cite{Becher:2006mr}:
\begin{eqnarray}
H(\hat{Q}^2,\mu_Q,\mu_T) &=& H(\hat{Q}^2,\mu_Q)\nonumber \\ &\times& \text{exp}\left\{ \int_{\mu_T^2}^{\mu_Q^2}  \frac{d \mu^2}{\mu^2} \left[ \text{ln}\frac{\hat{Q}^2}{\mu^2} \Gamma_H(\alpha_s (\mu)) +\gamma_H[\alpha_s (\mu)]\right] \right\},
\label{eq:hard_evolve}
\end{eqnarray}
where we have set $\hat{Q}^2 = x_1 x_2 Q^2$.  The expressions for the cusp anomalous dimension $\Gamma_H$ and $\gamma^V$ for the cases of Higgs and electroweak gauge boson production are given in Refs.~\cite{Becher:2006mr,Ahrens:2008nc}.  They have perturbative expansions, as do the $A$, $B$, and $C_a$ coefficients of the CSS approach.  A detailed study of the anomalous dimensions in SCET and in the standard QCD approach was previously studied in the context of threshold resummation~\cite{Becher:2006mr}.  The cusp anomalous dimension is equivalent to the $A$ factor appearing in the exponent to the order we are working, 
$\Gamma_H[\alpha_s (\mu)]=A(\alpha_s (\mu))$.  The leading terms in the anomalous dimensions that controls the single logarithm are the same: 
$\gamma^{V(1)} = B^{(1)}$.  At the two-loop level, this is no longer true; the effective theory organizes terms differently than the standard approach, and contributions from the matching coefficients $H(\hat{Q}^2,\mu_Q)$ and ${\cal G}^{ij}$: $\gamma^{V(2)} = B^{(2)}+$ contributions from $H(\hat{Q}^2,\mu_Q)$, ${\cal G}^{ij}$.  This has been observed in previous analyses comparing SCET evolution to the QCD literature~\cite{Idilbi:2005er,Becher:2006mr}.  A two-loop computation of ${\cal G}^{ij}$ is required to further check the relation between the CSS and effective theory approaches; this calculation is an important goal for future work.  By design, both approaches fully reproduce the low-$p_T$ limit of fixed-order result upon expansion in $\alpha_S$.  A check of the NLO $p_T$ spectrum would also require a two-loop computation of ${\cal G}^{qrs}$.

However, note that in the SCET approach, the low scale endpoint of the RG evolution of the Sudakov factor is at $\mu_T \sim p_T$. This differs from the standard approach where the corresponding endpoint is at $\mu \sim 1/b_\perp$ where $b_\perp$ is the impact parameter that is integrated over from zero to infinity. The limit of $b_\perp \to \infty$ gives rise to a Landau pole that must be dealt with by introducing an external prescription for any value of $p_T$. The SCET approach avoids this issue as the RG evolution is done entirely in momentum space.

 A well-known aspect of the CSS approach is its treatment of the limit $p_T \to 0$, $M\to \infty$.  It predicts that in this limit
$d\sigma / dp_T^2 $ goes like a power~\cite{Parisi:1979se} of $\Lambda_{QCD}/\hat{Q}$ and is thus sensitive to non-perturbative input. In the effective field theory approach this corresponds to the region where the TMF is no longer perturbative. The leading $1/p_T^2$ term coming from perturbative soft and collinear gluons is strongly Sudakov-suppressed by the evolution due to the cusp anomalous dimension in this limit.  The remaining contribution then comes from the non-perturbative region in the effective theory whose analysis remains to be done.

We also note that in our formalism, the factorization theorem is in terms of iBFs and the iSF which differ from the corresponding objects in the TMD factorization formalism ~\cite{Collins:1992kk, Collins:1999dz, Belitsky:2002sm, Ji:2004wu, Ji:2002aa, Hautmann:2007uw, Cherednikov:2007tw, Cherednikov:2008ua,Cherednikov:2009wk, Aybat:2011zv}. The iBFs and iSF depend on additional light-cone residual momentum components which regulate rapidity divergences as dictated by the physical kinematics of the process at finite $p_T$. This allows one to compute in perturbation theory the iBFs and iSF in pure dimensional regularization without the need for additional external regulators as in the TMD factorization approach. One can obtain a factorization formula in SCET analogous to the TMD factorization by expanding in these residual light cone momenta to get~\cite{Gao:2005iu}
\bea
\frac{d^2\sigma}{du \>dt} &=& \sum_{qijKL}\frac{\pi F^{KL;q}}{4Q^4N_c^2} \int d^2k_\perp \int \frac{d^2b_\perp}{(2\pi)^2} e^{i\vec{b}_\perp \cdot \vec{k}_\perp} \delta \Big [\omega_u \omega_t -\vec{k}_\perp^2 -M_z^2\Big ]\>H_Z^{KL;ijq}(\omega_u,\omega_t,\mu_Q;\mu_T)  \nn \\
&\times&J_n^{q}(\omega_u, 0,b_\perp,\mu_T) J_\bn^{\bar{q}}(\omega_t,0,b_\perp,\mu_T) S_{qq}(0,0,b_\perp,\mu_T),
\eea
where $\omega_u= \frac{M_z^2-u}{Q}, \omega_t=\frac{M_z^2-t}{Q}$.  The SCET objects $J_n^q, J_\bn^{\bar{q}},$ and $S_{qq}$ contain spurious rapidity divergences that require additional regulators beyond dimensional regularization.  For perturbative values of $p_T$, a matching calculation can be performed to write the above formula in terms of standard PDFs.  A more detailed comparison of our approach to the TMD factorization formalism is left for future work. 

\subsection{Expansion of resummed formula to higher order}

To demonstrate explicitly that our formalism correctly obtains the large logarithms of the CSS approach at higher orders, we expand the resummed $Z$-boson differential cross section of Eq.~(\ref{eq:anform}) to $\mathcal{O}(\alpha_s^2)$.  Our derivation closely follows the approach of Ref.~\cite{Altarelli:1984pt}.  We begin by inserting the partonic cross section of Eq.~(\ref{eq:anform}) into the hadronic convolution of Eq.~(\ref{hadcross}).  The resulting expression has the schematic form
\begin{equation}
\label{eq:hadcross2}
\frac{d^2 \sigma_Z}{dp_T^2 dY} = \int dx_1 dx_2 \, F(x_1,x_2) \,\delta \left( x_1 x_2 s + x_1 (t-M_Z^2) + x_2 (u-m_Z^2) +M_Z^2\right), 
\end{equation}
where $s$, $t$, and $u$ are the usual hadronic Mandelstam variables, defined for completeness in Appendix~\ref{appex-1}.  The function $F(x_1,x_2)$ denotes the contributions from the matrix elements and parton distribution functions.  The delta function comes from the partonic differential cross section, and allows one of the integrals over partonic momentum fractions to be immediately performed.  It is convenient to divide the integration in Eq.~(\ref{eq:hadcross2}) into two regions: one where the pseudorapidity of the parton recoiling against the $Z$ to give it a transverse momentum is greater than the $Z$ rapidity $Y$, and one where it is less than $Y$.  Doing so, and using the delta function to perform the $x_2$ integration in the first region and the $x_1$ integration in the second piece, leads to the expression
\begin{equation}
\frac{d^2 \sigma_Z}{dp_T^2 dY} = \int_{\sqrt{\tau_+}e^Y}^1 dx_1 \frac{F(x_1,x_2^{*})}{x_1 s+u-M_Z^2}
	+\int_{\sqrt{\tau_+}e^{-Y}}^1 dx_2 \frac{F(x_1^{*},x_2)}{x_2 s+t-M_Z^2},
\end{equation}
where 
\begin{eqnarray}
x_1^{*} &=& \frac{x_2 (M_Z^2-u)-M_Z^2}{x_2 s+t-M_Z^2},\nonumber \\
x_2^{*} &=& \frac{x_1 (M_Z^2-t)-M_Z^2}{x_1 s+u-M_Z^2},\nonumber \\
\sqrt{\tau_+} &=& \sqrt{\frac{p_T^2+M_Z^2}{s}}+\sqrt{\frac{p_T^2}{s}}.
\end{eqnarray}

To proceed, we now simplify the partonic cross section by expanding around the $p_T \to 0$ limit and keeping only the $1/p_T^2$ terms.  For simplicity we focus henceforth only on the $q\bar{q}$ partonic channel; the $qg$ channel proceeds identically.  In the first region of the integration, the partonic Mandelstam variables simplify in the $p_T \to 0$ limit as follows:
\begin{equation}
\hat{t} \to M_Z^2 \left( 1-\frac{x_1}{x_A}\right),\;\;\; \hat{u} \to 0,\;\;\; \hat{s} \to M_Z^2 \frac{x_1}{x_A},
\end{equation}
where we have introduced the notation
\begin{equation}
x_A = \frac{M_Z}{\sqrt{s}} e^Y, \;\;\; x_B= \frac{M_Z}{\sqrt{s}} e^{-Y}.
\end{equation}
The function $F$ appearing in the integrand takes the following form in the first region after this simplification:
\begin{equation}
F_{q\bar{q}}(x_1,x_2^{*}) \to f_{q/P}(x_1) f_{\bar{q}/P}(x_2^{*})  \times \frac{1}{p_T^2} \left[ 1+\left( \frac{x_A}{x_1}\right)^2\right],
\end{equation}
where for simplicity of presentation we have suppressed the overall constants which appear.  A similar simplification and structure are obtained in the other part of the integration.

We reduce this further by simplifying the remaining integrals over the $x_i$, following the procedure outlined in Ref.~\cite{Altarelli:1984pt}.  To facilitate comparison with results in the literature we introduce the standard notation for the convolution of two functions,
\begin{equation}
\left(f \otimes g \right) (z) = \int_0^1 dx dy \, f(x) g(y) \delta(z-xy),
\end{equation}
and remind the reader of the leading-order DGLAP kernel
\begin{equation}
P_{qq} (x) = C_F \left[ \frac{1+x^2}{1-x}\right]_+.
\end{equation}
We also introduce the following combinations of coupling constants to match the notation in Ref.~\cite{Arnold:1990yk}, with which we eventually compare:
\begin{equation}
e_{q\bar{q}}^2 = \frac{1}{16 {\rm cos}^2 \theta_W} \left[ 1+(1-4 |e_q| {\rm sin}^2 \theta_W)^2\right].
\end{equation}
For simplicity we continue to focus on the $q\bar{q}$ partonic channel.  After straightforward manipulations we arrive at our result for the differential distribution:
\begin{eqnarray}
\label{resummedresult}
\frac{d^2 \sigma_{Z,q\bar{q}}}{dp_T^2 dY} &=& \frac{4\pi^2}{3} \frac{\alpha}{{\rm sin}^2 \theta_W} e_{q\bar{q}}^2 \frac{\alpha_s(\mu_T)}{2\pi}\frac{1}{s \, p_T^2}
	\left \{ 2\,C_F f_{q/P}(x_A,\mu_T)  f_{\bar{q}/P}(x_B,\mu_T) \, {\rm ln} \frac{M_Z^2}{p_T^2} \right. \nonumber \\
	&-& 3\, C_F f_{q/P}(x_A,\mu_T)  f_{\bar{q}/P}(x_B,\mu_T)
	+ f_{q/P}(x_A,\mu_T) \left(P_{qq} \otimes f_{\bar{q}/P} \right)(x_B) \nonumber \\
	&+& \left. f_{\bar{q}/P}(x_B,\mu_T) \left(P_{qq} \otimes f_{q/P} \right)(x_A)
	\right \} \Big | {\rm exp}\left\{\frac{C_F}{4} \frac{\alpha_s}{\pi} \left[ -{\rm ln}^2 \frac{\mu_Q^2}{\mu_T^2} +3\,{\rm ln} \frac{\mu_Q^2}{\mu_T^2} \right] \right\} \Big |^2 . \nonumber \\
\end{eqnarray}
We have explicitly denoted the scales which appear in the overall coupling constant and in the PDFs.  We note that the solution for the evolution factor $U_{H_Z^q}$ can be obtained from Ref.~\cite{Becher:2006mr}; to the order in $\alpha_S$ we are working, the different momentum scales which appear in the evolution factors in the partonic cross section do not matter, and a simple overall exponential factor is obtained in the differential cross section.

To compare the structure of logarithms with those obtained in the CSS approach, we first use renormalization-group arguments to evolve all coupling constants which appear to an arbitrary renormalization scale $\mu_R$, and similarly use DGLAP to evolve all PDFs to the factorization scale $\mu_F$.  We organize our result following the notation of Ref.~\cite{Arnold:1990yk} into a joint expansion in $\alpha_s$ and ${\rm ln}(M_Z / p_T)$:
\begin{equation}
\frac{d^2 \sigma_{Z,q\bar{q}}}{dp_T^2 dY} = \frac{4\pi^2}{3} \frac{\alpha}{{\rm sin}^2 \theta_W} e_{q\bar{q}}^2 \, \frac{1}{s \, p_T^2}
\sum_{m,n}  \left( \frac{\alpha_s(\mu_R)}{2\pi}\right)^n {_{n}D_{m}} \,{\rm ln}^m \frac{M_Z^2}{p_T^2}.
\end{equation}
We set $\mu_Q = M_Z$ and $\mu_T = p_T$ (we comment later on the choice of an imaginary matching scale $\mu_Q$, as suggested recently~\cite{Ahrens:2008qu}).  Only terms through ${\cal O}(\alpha_s^2)$ are kept.  We introduce the explicit forms for the first few coefficients appearing in the CSS expansion of Eq.~(\ref{csscoeffs}): $A^{(1)} = 2\, C_F$, $B^{(1)} = -3\, C_F$. Introducing the nomenclature $f_{q/P}(x_A,\mu_F) = f_A$, $f_{{\bar q}/P}(x_B,\mu_F) = f_B$, we find the following results for the first few coefficients:
\begin{eqnarray}
{_{1}D_{1}} &=& A^{(1)} f_A f_B, \nonumber \\
{_{1}D_{0}} &=& B^{(1)} f_A f_B + f_B \left( P_{qq} \otimes f \right)_A + f_A \left( P_{qq} \otimes f \right)_B , \nonumber \\
{_{2}D_{3}} &=& -\frac{1}{2} \left[ A^{(1)}\right]^2 f_A f_B, \nonumber \\
{_{2}D_{2}} &=& -\frac{3}{2}A^{(1)} \left[ f_B \left( P_{qq} \otimes f \right)_A + f_A \left( P_{qq} \otimes f \right)_B \right] 
	-\left[ \frac{3}{2} A^{(1)} B^{(1)} - \beta_0 A^{(1)}\right] f_A f_B, \nonumber \\
{_{2}D_{1}} &=& \left\{ -A^{(1)} f_B \left( P_{qq} \otimes f \right)_A \text{ln} \frac{\mu_F^2}{M_Z^2}-2 B^{(1)} f_B \left( P_{qq} \otimes f \right)_A -\frac{1}{2} \left[ B^{(1)} \right]^2 f_A f_B \right. \nonumber \\
	&& +\frac{\beta_0}{2} A^{(1)} f_A f_B \,{\rm ln} \frac{\mu_R^2}{M_Z^2}  +\frac{\beta_0}{2} B^{(1)} f_A f_B - \left( P_{qq} \otimes f \right)_A\left( P_{qq} 	\otimes f \right)_B \nonumber \\
	&& \left. - f_B \left( P_{qq} \otimes P_{qq} \otimes f \right)_A + \beta_0\, f_B  \left( P_{qq} \otimes f \right)_A \right\} + \left[ A \leftrightarrow B \right].
\end{eqnarray}
The coefficients ${_{1}D_{1}}$, ${_{1}D_{0}}$, ${_{2}D_{3}}$, and ${_{2}D_{2}}$ agree\footnote{We disagree with the statement made in Ref.~\cite{Becher:2010tm} that our formalism does not correctly resum logarithms  at the next-to-leading-logarithmic order; our explicit check makes it clear that this claim is incorrect.} with the analogous ${_{n}C_{m}}$ coefficients of Ref.~\cite{Arnold:1990yk} that appear in both the fixed-order expansion and the CSS formalism.  Differences occur in ${_{2}D_{1}}$; the ${_{2}C_{1}}$
formalism of the usual approach contains two additional terms depending on the quantities $A^{(2)}$ and $C^{(1)}$.  This is not surprising, as our result has been computed  only to next-to-leading logarithmic accuracy.  These terms in the expansion are of next-to-next-to-leading logarithmic order. Denoting $L = {\rm ln} \, M_Z^2/p_T^2$, we remind the reader that resummation to a given order gives the following towers of logarithms~\cite{Ellis:1997ii}:
\begin{eqnarray}
\text{leading logarithmic} &:& \alpha_s^{n}L^{2n-1}, \nonumber \\
\text{next-to-leading logarithmic} &:& \alpha_s^{n}L^{2n-2}, \nonumber \\
\text{next-to-next-to-leading logarithmic} &:& \alpha_s^{n}L^{2n-3}.
\end{eqnarray}
The full result at next-to-next-leading logarithmic accuracy  along with the complete result for ${_{2}D_{1}}$ requires the next higher order calculation of the TMF.  However,  some of the next-to-next-to-leading order logarithmic terms can be already seen to appear in the partial result for ${_{2}D_{1}}$ above.  In our factorization formula, we obtain two distinct large logarithms: explicit logarithms ${\rm ln} \, \mu_Q^2/\mu_T^2$ coming from the resummed exponent, and  a kinematic one of the form ${\rm ln} \, M_Z^2/p_T^2$ coming from the hadronic convolution.  This organization is different than in the CSS approach, but all required terms are correctly obtained.

It was recently suggested in the literature to utilize an imaginary matching scale for $\mu_Q$, which has the effect of resumming factors of $\pi^2$ which arise from the time-like momentum transfer~\cite{Ahrens:2008qu, Ahrens:2008nc}.  This was shown to improve the convergence of the perturbative expansion for inclusive Higgs production \cite{Ahrens:2008qu, Ahrens:2008nc}, and has also been utilized in the literature to study Drell-Yan~\cite{Stewart:2010pd}.  Doing so here adds the following additional term at the single-logarithmic order:
\begin{equation}
{_{2}D_{1}} \to {_{2}D_{1}} + C_F\, \pi^2 f_A f_B.
\end{equation}
This factor is part of the contribution to the coefficient $C^{(1)}$ in the CSS approach.

\section{Numerical Results}
\label{sec:numbers}

We present here numerical predictions utilizing the factorization and resummation formulae we have derived.  We show results for Higgs production at a 7 TeV LHC, and for $Z$ production at the Tevatron. Our numerical results are based on the resummed partonic cross sections presented in Eq.~(\ref{eq:anform}).  For the $Z$ we compare with the Run 1 data from both CDF and D0 to demonstrate the consistency of our calculation with experimental results. Our results are model independent and free of Landau pole prescriptions required in the standard approach.  For perturbative values of $p_T$, the transverse momentum distribution is given entirely in terms of field-theoretically derived perturbative functions and the standard initial state PDFs. For non-perturbative values of $p_T$, the TMF function ${\cal G}$ is non-perturbative, but field-theoretically well-defined, and can be modeled and extracted from data. In this section, we restrict our results only to perturbative values of $p_T$ leaving the non-perturbative region for future work.

Before describing the parameter values assumed in our study, we comment on the order of our resummation.  The hard matching coefficient $H$, the cusp and non-cusp anomalous dimensions $\Gamma_H$ and $\gamma_H$ respectively, and the TMF $\mathcal{G}$ all have perturbative expansions that must be calculated to sufficiently high order to achieve resummation of certain classes of logarithms.  
We note that since generating a finite $p_T$ requires the emission of at least one parton, the contribution of the TMF to the transverse momentum spectrum begins only at 1-loop.  All quantities required to achieve next-to-leading logarithmic accuracy (NLL) are known, and have been detailed in previous sections of this paper.  To achieve NNLL precision, the 2-loop result for $\mathcal{G}$ is needed.  


\begin{figure}[h]
\includegraphics[angle=90,scale=0.6]{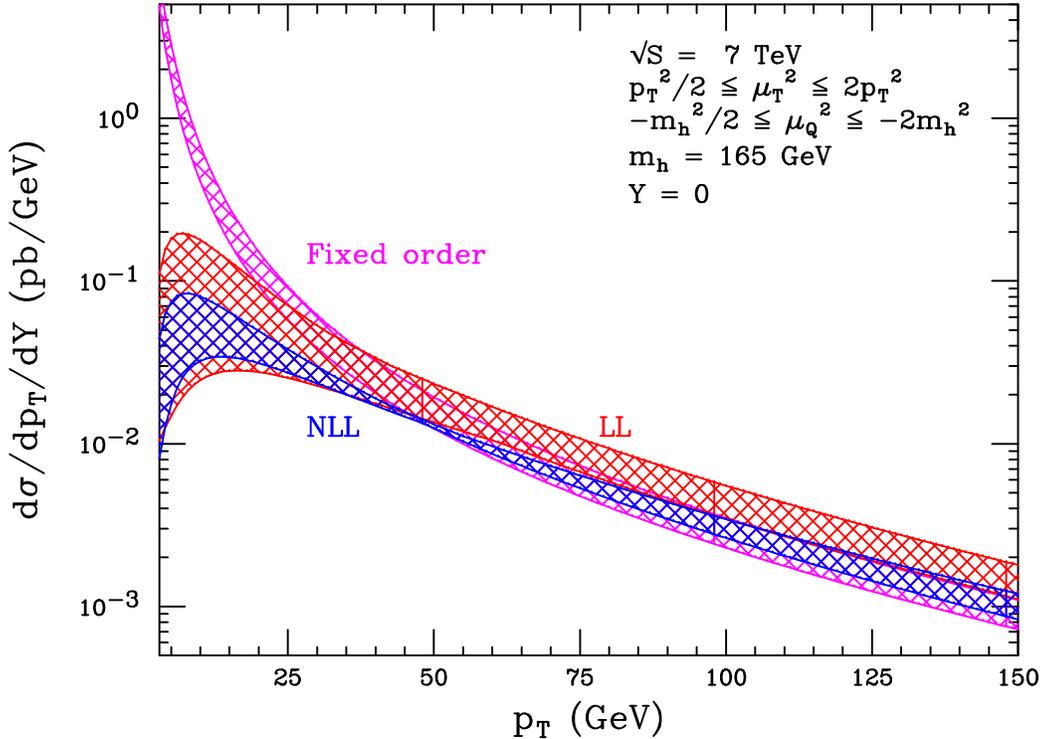}
\caption{Numerical predictions for the transverse momentum spectrum for Higgs boson production at the LHC for central rapidity.  Shown are the fixed-order result and those obtained after implementing the resummation formula of Eq.~(\ref{intro-DY}) through LL and NLL.  The bands arise from the scale variation shown in the text.}
\label{higgsplot}
\end{figure}

We start by discussing some general features of our numerical results.  We show fixed-order results at leading order in perturbation theory, and results at both LL and NLL matched to the fixed-order results at ${\cal O}(\alpha_s)$, as shown in Section~\ref{sec:analytic}.  In the standard nomenclature these would be called LL+LO and NLL+LO.  We use MSTW 2008 parton distribution functions~\cite{Martin:2009iq}.  
\begin{figure}
\includegraphics[angle=90,scale=0.6]{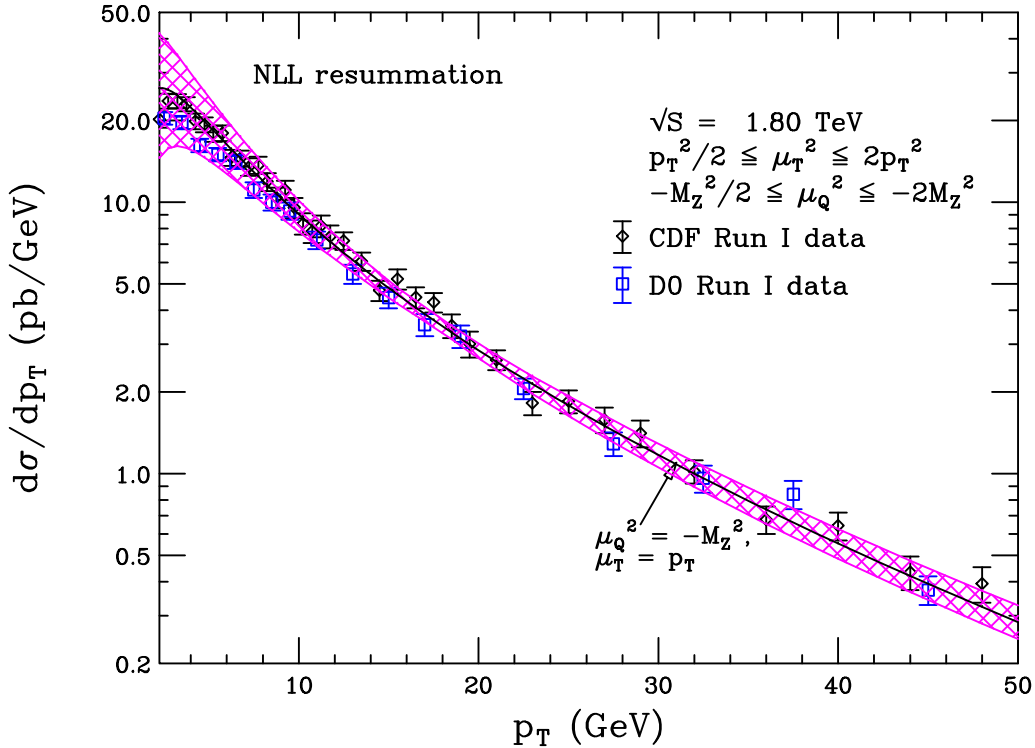}
\caption{Numerical predictions for the transverse momentum spectrum for $Z$ boson production at Tevatron Run 1, compared with data form both CDF and D0.  Shown is the resummation prediction of Eq.~(\ref{intro-G}) at NLL.  The bands arise from the scale variation shown in the text, while the result for the central scale choice is shown by the solid line.  The lower limit of the plot is $p_T$= 1.75 GeV.}
\label{Zplot}
\end{figure}
For LL and LO predictions we use leading order PDFs with 1-loop running of the strong coupling constant, while for our NLL results we use NLO PDFs with 2-loop running for $\alpha_s$.  Our results depend on the two matching scales $\mu_T$ and $\mu_Q$.  The dependence of the cross section on these arbitrary scales occurs at one order beyond the order in $\alpha_s$ to 
which we have calculated; it would vanish completely if we could compute the cross section to all orders in perturbation theory.  The variation of these scales therefore provides some 
indication of missing higher-order effects, and is conventionally used as an estimate of the theoretical uncertainty.  We must choose both a central value for these scales and a range of 
variation to obtain an uncertainty estimate.  As our central scale choices we set 
$\mu_T^2 = p_T^2$ and $\mu_Q^2= -M^2$.  These are chosen to minimize logarithms that appear in the perturbative expansions of the hard function and the TMF, as discussed in Section~\ref{sec:review}.  We vary $\mu_T^2$, $\mu_Q^2$ independently around these choices by a factor of 2.  Two unconventional aspects of these choices require comment.  Following Ref.~\cite{Ahrens:2008qu}, we utilize an imaginary matching scale for $\mu_Q$, which has the effect of resumming factors of $\pi^2$ which arise from the time-like momentum transfer appearing in $H$.  
This was shown to improve the convergence of the perturbative expansion for inclusive Higgs production \cite{Ahrens:2008qu, Ahrens:2008nc}, and has also been utilized in the literature to study Drell-Yan~\cite{Stewart:2010pd}. We also find better agreement with data (see Fig.~\ref{Zplot})  for an imaginary $\mu_Q$ compared to a real $\mu_Q$ which can be attributed to the effect of resumming factors of $\pi^2$ with the former choice.    We also choose to vary our scales around a reduced range to avoid evaluating $\alpha_s(\mu_T)$ at a non-perturbative scale when the transverse momentum becomes small.  An framework for incorporating the non-perturbative region of transverse momentum into the SCET formalism was given in 
Ref.~\cite{Mantry:2010bi}.  In this approach the scale $\mu_T$ freezes at a value near $\Lambda_{QCD}$ as the $p_T$ approaches zero.

In Fig.~\ref{higgsplot} we show the predictions for the Higgs $p_T$ spectrum at the LHC, using both the fixed-order expression and the resummed results at LL and NLL accuracies.  The general features of this plot are clear: large logarithms of the form $\text{ln}\, (m_h^2/p_T^2)$ spoil the fixed-order perturbative expansion at low $p_T$.  The Sudakov suppression coming from the renormalization-group evolution of the hard function $H$ tames this behavior.  The central value of the prediction is absolutely stable upon proceeding from LL to NLL; only a reduction of the scale variation is observed.  At intermediate and high momenta, the matching onto the fixed-order expression is smooth.  The sensitivity to scale choices that can lead to negative results~\cite{Bozzi:2005wk} in the standard approach, does not occur in this effective field theory approach.  This allows the matching scale $\mu_Q$ to be varied throughout a range sufficient to use it as an estimator of the theoretical uncertainty.  An additional error also arises from imprecise knowledge of parton distribution functions.  We postpone a numerical analysis of this issue 
until a detailed study of boson $p_T$ distributions at both the Tevatron and LHC incorporating the non-perturbative region is performed.

One aspect of transverse resummation in SCET that requires further study is the treatment of the non-perturbative region $p_T \sim \Lambda_{QCD}$.  In our analysis, the transverse momentum function $\mathcal{G}$ becomes non-perturbative, and must be modeled.  The onset of this region can be seen in the plot by the large scale variation at low $p_T$, which is caused by evaluating $\alpha_s(\mu_T) \sim \alpha_s(\Lambda_{QCD})$.  Since this object has a non-pertubative definition in terms of operator matrix elements and a well-defined running, it is reasonable to extract this function using available data.  In our plot for the Higgs $p_T$ distribution we simply stop our plot at a lower value of $p_T = 3$ GeV.  The study of the 
non-perturbative region of $p_T$ was recently begun in Ref.~\cite{Mantry:2010bi}.

In Fig.~\ref{Zplot} we plot our prediction for the $Z$-boson $p_T$ distribution at the Tevatron Run 1, and compare to data from CDF~\cite{Affolder:1999jh} and D0~\cite{Abbott:1999yd}.  We study the spectrum down to $p_T=1.75$ GeV.  The agreement with the data is excellent over the entire range.  The low $p_T$ version of this data can eventually be used to constrain the non-perturbative TMF that appears in SCET, as is done in the CSS approach~\cite{Landry:1999an}.

\section{Conclusions}
\label{sec:conc}

In this manuscript we have extended our analysis of transverse momentum distributions using the Soft-Collinear Effective Theory(SCET) to account for both electroweak and Higgs boson production at low $p_T$ in hadronic collisions.  We have derived a factorization theorem for the transverse momentum distribution for the production of electroweak gauge boson production, and have provided all necessary analytic expressions to perform resummation of low-$p_T$ distributions for any color-neutral particles to next-to-leading-logarithmic accuracy.  Our effective field theory approach is free of  the Landau pole that appears in the standard approach even for perturbative values of $p_T$. We thus have a numerically stable matching to the fixed-order QCD result, leading to a smooth transition from the low-$p_T$ resummation  region to the intermediate and high $p_T$ region without the need for a matching prescription. For perturbative values of $p_T$, our approach predicts the transverse momentum distribution entirely in terms of field-theoretically derived perturbative functions and standard initial state PDFs. For non-perturbative values of $p_T$, an additional non-perturbative Transverse Momentum Function (TMF) appears with a rigorous field-theoretic definition and computable anomalous dimension.

 We have presented the first numerical predictions for $p_T$ spectra arising from SCET for Higgs and $Z$-boson production, and for $Z$ boson production have shown an initial numerical comparison with Tevatron data.  The agreement with data is excellent over the kinematic range currently covered by our factorization formula, indicating that SCET will provide a useful framework for the analysis and interpretation of hadron collider distributions.

Our analysis reveals several directions for future work.  Precision predictions at next-to-next-to-leading logarithmic accuracy require  two-loop computations of the iBFs and iSFs that appear in our factorization theorems.  This computation is within current technical capabilities. The region of non-perturbative $p_T$ requires further study through a modeling of the non-perturbative TMF followed by its extraction from data.

In summary, the SCET approach to transverse momentum resummation offers a compelling alternative to the usual method.  Several explicit checks have been performed: (1) a comparison with the leading fixed order cross-section, (2) the cancellation of the scale dependence between the various components in the factorization theorem as determined by their RG evolution structure, and (3) an explicit check of the logarithms at next-to-leading logarithmic order. Furthermore, we find good agreement with data. We look forward to the further development of our results.

\section*{Acknowledgments}

\noindent
This work was supported by the DOE grants DE-FG02-95ER40896 and DE-FG02-08ER4153.

\appendix

\section{Factorization of electroweak gauge boson differential distributions}
\label{appex-1}

In this appendix we describe the steps in the derivation of the factorization formula for  the transverse momentum and rapidity distributions
for electroweak gauge boson production.  These steps closely follow the derivation of the analogous factorization formula for Higgs production derived 
in~\cite{Mantry:2009qz}. 
For simplicity in notation we focus here on the case of single-boson production. With straightforward modifications,
one can obtain analogous factorization formulae for neutral-current $\gamma^* +Z$ production as well as the case where the final state leptonic
decay products of the vector boson are treated differentially. It is convenient to work with the hadronic Mandelstam variables $u,t$ which are related to $p_T$ and $Y$ as 
\bea
\label{mandel-1}
u = (p_2 -q)^2 = M_Z^2 - Q \sqrt{p_T^2 + M_Z^2} e^Y, \nn \\
t= (p_1 - q)^2 = M_Z^2 - Q \sqrt{p_T^2 + M_Z^2} e^{-Y}, \nn \\
\eea
where $q^\mu$ and $M_Z$ denote the  vector-boson momentum and mass respectively and\footnote{Note that $Q=\bn \cdot p_1 = n\cdot p_2$ denotes the hadronic center of mass energy, and is not related to $q^\mu$ which is the vector-boson momentum and satisfies $q^2=M_Z^2$.}   
\bea
\label{mandel-2}
du\> dt &=& Q^2 dp_T^2 dY.
\eea

After matching the vector-boson production current onto SCET$_{p_T}$ current as explained in section \ref{sec:hard}, the differential cross-section in the hadronic Mandelstam variables takes the form
\bea
\frac{d^2\sigma}{du\> dt}&=& \frac{1}{2Q^2}\Big [ \frac{1}{4}\Big ] \int \frac{d^4q}{(2\pi)^4}(2\pi) \theta(q^0) \> \delta (q^2 - M_Z^2)  L^{\mu \nu}(q) \int d\omega_1 \int d\omega_2 \int d\omega_1' \int d\omega_2' \nn \\
&\times&\sum_{qq'ijKL}  \sum_{X_n,X_\bn,X_{s}} (2\pi)^4 \delta^{(4)}(p_1+p_2 - q - P_{X_n}-P_{X_\bn}-P_{X_{s}})\delta \big [u - (p_2-q)^2 \big ]\delta \big [t - (p_1-q)^2 \big ]    \nn \\
&\times&   C^{K;iq}(\omega_1,\omega_2)C^{*L;jq'}(\omega_1',\omega_2') \langle p_1 p_2 | {\cal O}^{Lq'\dagger}_\nu (\omega_1',\omega_2') (0) | X_n,X_\bn,X_{s} \rangle  \nn\\
&\times& \langle X_n,X_\bn,X_{s} | {\cal O}^{Kq}_\mu (\omega_1,\omega_2) (0) |p_1p_2\rangle, \nn \\
\eea
where  the indices run over
\bea
K,L &=&\{V,A\}, \qquad q,q',i,j = \{u,d,s,\cdots \},
\eea
where $V$ and $A$ label the vector and axial-vector Dirac structure. The indices $q,q',i,j$ run over the massless quarks that appear in the initial protons. The contribution from a pure gluon SCET$_{p_T}$ operator that would produce $gg \to V$ with $V={\gamma^*,Z,W}$ vanishes for Drell-Yan processes, so that the sum over $i,j$ does not include the gluon. The
 overall factor of $\frac{1}{4}$ comes from averaging over the initial hadron spins, the final state $|X\rangle$ has been broken up into the $n$-collinear, $\bn$-collinear, and soft states so that $|X\rangle = |X_n X_\bn X_s \rangle$, and the SCET operators have the form
\bea
{\cal O}_\mu^{Kq} (\omega_1,\omega_2) &\equiv & (  \bar{\xi}_q W  )_{\bn,\omega_2} S_\bn^\dagger \Gamma_\mu^{Kq} S_n ( W^\dagger \xi_q )_{n,\omega_1}, \qquad {\cal O}^{Lq'\dagger}_\nu (\omega_1',\omega_2') \equiv  (  \bar{\xi}_{q'} W  )_{n,\omega_1'} S_n^\dagger \Gamma_\nu^{Lq'\dagger} S_\bn( W^\dagger \xi_{q'} )_{\bn,\omega_2'}, \nn \\
\eea
where the Dirac structures $\Gamma_\mu^{Kq}$  are given by
\bea
\Gamma_\mu^{Vq} &=& g_V^q\gamma_\mu^\perp , \qquad \Gamma_\mu^{Aq}= g_A^q \gamma_\mu^\perp \gamma_5.
\eea
$g_V^q$ and $g_A^q$ denote the vector and axial-vector couplings of the q-th quark to the  vector boson respectively. The tensor $L^{\mu \nu}$ denotes the product of the leptonic currents arising from the vector-boson decay.  For simplicity of notation, we will present out formulae integrated over the leptonic phase space, so that we can use effective polarization vectors and set
\bea
\label{vec-pol}
L^{\mu \nu}(q) &=& \sum_{\text{pols.}} \epsilon^\mu(q) \epsilon^{*\nu}(q) = -g^{\mu \nu} + \frac{q^\mu q^\nu}{M_Z^2}.
\eea
By the equations of motion for massless quarks, the contribution of the $q^\mu q^\nu$ term in the above polarization sum vanishes when
contracted with the quark bilinear currents. This allows us to effectively set $L_V^{\mu \nu}(q) \to -g^{\mu \nu}$ which is used in the rest of this analysis.
Next we use the soft-collinear  decoupling~\cite{Bauer:2000yr, Bauer:2001yt} property of the leading order SCET$_{p_T}$ Lagrangian
to decouple the matrix elements into $n$-collinear, $\bn$-collinear, and soft objects so that the differential cross-section becomes
\bea
\frac{d^2\sigma}{du\> dt}&=&\frac{-g^{\mu \nu}}{8Q^2} \int \frac{d^4q}{(2\pi)^3} \theta(q^0) \> \delta (q^2 - M_Z^2) \delta \big [u - (p_2-q)^2 \big ]\delta \big [t - (p_1-q)^2 \big ]     \int d\omega_2 \int d\omega_1' \int d\omega_2' \nn \\
&\times& \sum_{qq'ijKL} \sum_{X_n,X_\bn,X_{s}} (2\pi)^4 \delta^{(4)}(p_1+p_2 - q - P_{X_n} -P_{X_\bn} - P_{X_{s}})  \nn \\
&\times& C^{K;iq}(\omega_1,\omega_2)C^{*L;jq'}(\omega_1',\omega_2')(\Gamma_\mu^{Kq})_{\omega \sigma} (\Gamma_\nu^{Lq'\dagger})_{\alpha \beta}\langle p_1 | (\bar{\xi}_{q'} W)_{n,\omega_1'}^{\alpha a}(0) | X_n \rangle  \langle p_2 | (W^\dagger \xi_{q'})_{\bn,\omega_2'}^{\beta b}(0) | X_\bn \rangle\nn \\
&\times&  \langle X_n | (W^\dagger \xi_q)^{\sigma d}_{n,\omega_1}(0)| p_1\rangle \langle X_\bn | (\bar{\xi}_q W)_{\bn,\omega_2}^{\omega e}(0) |p_2 \rangle  \langle 0 |(S_n^\dagger)^{ac} (S_\bn)^{cb} |X_{s} \rangle \langle X_{s} | (S_\bn^\dagger)^{ef} (S_n)^{fd} |0\rangle .\nn \\
\eea
We perform a series of steps that allow us perform the sum over the states $X_n,X_\bn, X_s$ while consistently maintaining the final state restriction on the gauge-boson momentum. We begin by inserting the identity operator
\bea
\label{identity}
1 &=& \int d^4p_n \int d^4p_\bn \int d^4p_{s} \delta^{(4)}(p_n-P_{X_n})\delta^{(4)}(p_\bn-P_{X_\bn}) \delta^{(4)}(p_{s}-P_{X_{s}}),
\eea
and decompose the momenta into label and residual parts so that
\bea
P_{X_n}^- = \tilde{P}_{X_n}^- + K_{X_n}^-, \qquad P_{X_\bn}^+=\tilde{P}_{X_\bn}^+ + K_{X_\bn}^+, 
\eea
where $ \tilde{P}_{X_n}^-, \tilde{P}_{X_\bn}^+ \sim M_Z$ and $K_{X_n}^-,  K_{X_\bn}^+ \ll M_Z$.  We write the remaining momentum
components as 
\bea
P_{X_n}^{+,\perp} = K_{X_n}^{-,\perp}, \qquad P_{X_\bn}^{-,\perp} = K_{X_\bn}^{-,\perp},   \qquad P_{X_{s}} = K_{X_{s}}
\eea
and similarly write
\bea
p_n^- = \tilde{p}_n^- + k_n^-, \qquad p_\bn^+ = \tilde{p}_\bn^+ + k_\bn^+, \nn \\
p_n^{+,\perp} = k_n^{+,\perp}, \qquad p_\bn^{-,\perp} = k_\bn^{-,\perp} \qquad p_{s}^\mu = k_{s}^\mu,
\eea
where again $\tilde{p}_n^-,  \tilde{p}_\bn^+ \sim M_Z$ and $k_n^-,k_\bn^+ \ll M_Z$. The delta functions in Eq.~(\ref{identity}) can be broken up into Kronecker deltas over label momenta and residual delta functions which we can write using the integral representation as
\bea
\label{identity-1}
1 &=& \sum_{\tilde{p}_n^-,\tilde{p}_\bn^+} \delta_{\tilde{p}_n^-,\tilde{P}_{X_n}^-}\delta_{\tilde{p}_\bn^+,\tilde{P}_{X_\bn}^+} \int d^4k_n d^4k_\bn d^4k_{s} \> \delta^{(4)}(k_n-K_{X_n})\delta^{(4)}(k_\bn-K_{X_\bn})\delta^{(4)}(k_{s}-K_{X_{s}}) \nn \\
&=& \sum_{\tilde{p}_n^-,\tilde{p}_\bn^+} \delta_{\tilde{p}_n^-,\tilde{P}_{X_n}^-}\delta_{\tilde{p}_\bn^+,\tilde{P}_{X_\bn}^+} \int d^4k_n d^4k_\bn d^4k_{s} \int \frac{d^4x}{(2\pi)^4}\int \frac{d^4y}{(2\pi)^4}\int \frac{d^4z}{(2\pi)^4} e^{i(k_n - K_{X_n})\cdot x} e^{i(k_\bn - K_{X_\bn})\cdot y} e^{i(k_{s} - K_{X_{s}})\cdot z}. \nn \\
\eea
Similarly, the momentum of the vector boson  can be divided into label and residual components so that
\bea 
n\cdot q &=& n \cdot \tilde{q} + n \cdot k, \qquad
\bn\cdot q = \bn \cdot \tilde{q} + \bn \cdot k, \qquad
\vec{q}_\perp = \vec{k}_\perp, \nn \\
\eea 
where the $n\cdot q, \bn \cdot q \sim M_Z$ and $n\cdot k, \bn \cdot k, k_\perp \ll M_Z$. The phase space integral over $q^\mu$ can now we written as
\bea
\label{q-phase-1}
\int d^4q \> \delta (q^2 - M_Z^2) &=& \sum_{\tilde{q}^+,\tilde{q}^-} \int d^2k_\perp \int \frac{dk^+ dk^-}{2}
  \delta (\tilde{q}^+ \tilde{q}^- + \tilde{q}^+k^- + \tilde{q}^-k^+ + k^+k^- -\vec{k}_\perp^2 - M_Z^2).\nn \\
\eea
The four-momentum conserving delta function appearing in the differential cross-section can be written as a product of Kronecker delta functions over label momenta and delta functions over residual momenta as
\bea
\label{delta-label-res}
\delta^{(4)}(p_1+p_2 - q - P_{X_n} -P_{X_\bn} - P_{X_{s}}) &=& \delta_{\omega_1,\tilde{q}^-} \delta_{\omega_2,\tilde{q}^+} \delta^{(2)} (K_{X_{s}}^\perp + K_{X_n}^\perp + K_{X_\bn}^\perp + k_\perp)\nn \\
&\times& \delta (K_{X_n}^+ + K_{X_\bn}^+ + K_{X_{s}}^+ + k^+) \delta (K_{X_n}^- + K_{X_\bn}^- + K_{X_{s}}^- + k^-) \nn \\
&=& \delta_{\omega_1,\tilde{q}^-} \delta_{\omega_2,\tilde{q}^+}  \delta^{(2)} (k_{s}^\perp + k_n^\perp + k_\bn^\perp + k_\perp)\nn \\
&\times& \delta (k_n^+ + k_\bn^+ + k_{s}^+ + k^+) \delta (k_n^- + k_\bn^- + k_{s}^- + k^-), \nn \\
\eea
where we used the residual delta functions in the first line of Eq.(\ref{identity-1}) to obtain the second equality above. Using Eqs. (\ref{identity-1}), (\ref{q-phase-1}), and (\ref{delta-label-res}) the cross-section can be brought into the form
\bea
\frac{d^2\sigma}{du\> dt}&=& \frac{-\pi g^{\mu \nu}}{4 Q^2N_c^2}\sum_{qq'ijKL}\sum_{\tilde{q}^+,\tilde{q}^-} \int d^2k_\perp \int \frac{dk^+ dk^-}{2} \delta (\tilde{q}^+ \tilde{q}^- + \tilde{q}^+k^- + \tilde{q}^-k^+ + k^+k^- - M_Z^2) \nn \\
&\times& \int d\omega_1 \int d\omega_2\>  C^{K;iq}(\omega_1,\omega_2)C^{*L;jq'}(\omega_1,\omega_2)\int d^4k_n d^4k_\bn d^4k_{s} \int \frac{d^4x}{(2\pi)^4}\int \frac{d^4y}{(2\pi)^4}\int \frac{d^4z}{(2\pi)^4} \nn \\
&\times& e^{ik_n\cdot x} e^{ik_\bn\cdot y} e^{ik_{s} \cdot z} \delta_{\omega_1,\tilde{q}^-} \delta_{\omega_2,\tilde{q}^+}\>\delta^{(2)} (k_{s}^\perp + k_n^\perp + k_\bn^\perp + k_\perp) \nn \\
&\times& \delta (k_n^+ + k_\bn^+ + k_{s}^+ + k^+) \delta (k_n^- + k_\bn^- + k_{s}^- + k^-)\nn \\
&\times& \delta \big [u - M_Z^2 +  Q \>\bn \cdot q \big ] \delta \big [t - M_Z^2 +  Q \>n \cdot q \big ]    \nn \\
&\times& \delta_{qq'}\delta_{qq'} (\Gamma_\mu^{Kq})_{\omega \sigma} (\Gamma_\nu^{Lq'\dagger})_{\alpha \beta} \langle p_1 | (\bar{\xi}_{q'} W)_{n}^{\alpha a}(x)  (W^\dagger \xi_q)^{\sigma a}_{n,\omega_1}(0)| p_1\rangle  \langle p_2 | (W^\dagger \xi_{q'})_{\bn}^{\beta b}(y) (\bar{\xi}_q W)_{\bn,\omega_2}^{\omega b}(0) |p_2 \rangle   \nn \\
&\times& \text{Tr}\> \langle 0 |\bar{T}[(S_n^\dagger)^{ac} (S_\bn)^{cb}(z) ] T[(S_\bn^\dagger)^{ef} (S_n)^{fd} ]|0\rangle ,
\eea
where we have simplified the color structure using the identities
\bea
 \langle p_1 | (\bar{\xi}_{q'} W)_{n}^{\alpha a}(x) | X_n \rangle \langle X_n | (W^\dagger \xi_q)^{\sigma d}_{n,\omega_1}(0)| p_1\rangle &=&  \frac{\delta^{qq'}\delta^{ad}}{N_c}  \langle p_1 | (\bar{\xi}_j W)_{n}^{\alpha f}(x) | X_n \rangle \langle X_n | (W^\dagger \xi_i)^{\sigma f}_{n,\omega_1}(0)| p_1\rangle  \nn \\
  \langle p_2 | (W^\dagger \xi_{q'})_{\bn}^{\beta b}(y) | X_\bn \rangle \langle X_\bn | (\bar{\xi}_q W)_{\bn,\omega_2}^{\omega e}(0) |p_2 \rangle &=& \frac{\delta^{qq'}\delta^{be}}{N_c}   \langle p_2 | (W^\dagger \xi_j)_{\bn}^{\beta f}(y) | X_\bn \rangle \langle X_\bn | (\bar{\xi}_i W)_{\bn,\omega_2}^{\omega f}(0) |p_2 \rangle. \nn \\
\eea
Next we apply a spin Fierz identity which allows us to bring the cross-section into the form
\bea
\label{cross-section-3}
\frac{d^2\sigma}{du\> dt}&=& \sum _{qijKL} \frac{\pi F^{KL;q}}{4Q^2N_c^2} \sum_{\tilde{q}^+,\tilde{q}^-} \int d^2k_\perp \int \frac{dk^+ dk^-}{2} \delta (\tilde{q}^+ \tilde{q}^- + \tilde{q}^+k^- + \tilde{q}^-k^+ + k^+k^- - M_Z^2) \nn \\
&\times& \int d\omega_1 \int d\omega_2  \>H_Z^{KL;ijq}(\omega_1,\omega_2,\mu_Q;\mu_T) \int d^4k_n d^4k_\bn d^4k_{s} \int \frac{d^4x}{(2\pi)^4}\int \frac{d^4y}{(2\pi)^4}\int \frac{d^4z}{(2\pi)^4} \nn \\
&\times& e^{ik_n\cdot x} e^{ik_\bn\cdot y} e^{ik_{s} \cdot z} \delta_{\omega_1,\tilde{q}^-} \delta_{\omega_2,\tilde{q}^+} \delta^{(2)} (k_{s}^\perp + k_n^\perp + k_\bn^\perp + k_\perp) \delta (k_n^+ + k_\bn^+ + k_{s}^+ + k^+)\nn \\
&\times& \delta (k_n^- + k_\bn^- + k_{s}^- + k^-) \delta \big [u - M_Z^2 +  Q \>\bn \cdot q \big ] \delta \big [t - M_Z^2 +  Q \>n \cdot q \big ] \nn \\
&\times&J_n^{q}(\omega_1, x, \mu_T) J_\bn^{q}(\omega_2, y,\mu_T)S_{qq}(z,\mu_T), 
\eea
where we have defined the hard function
\bea
\label{hard-1}
H_Z^{KL;ijq}(\omega_1,\omega_2,\mu) = C^{K;iq}(\omega_1,\omega_2,\mu)C^{*L;jq}(\omega_1,\omega_2,\mu),
\eea
 and $H_Z^{KL;ijq}(\omega_1,\omega_2,\mu_Q;\mu_T)$ denotes the RG-evolved hard function from the scale $\mu_Q \sim M_Z$ to $\mu_T \sim p_T$.   The quantity $F^{KL;i}$ comes form the contraction of the leptonic tensor with the Dirac structure of the hadronic tensor, and is given by
\bea
\label{FKL}
F^{KL;q} &=& -g^{\mu \nu}\frac{\text{Tr}\>[\nslash \Gamma_\mu^{Kq} \bnslash \Gamma_\nu^{Lq\dagger}]}{16} 
\eea
The jet and soft functions are given by
\bea
J_n^{q}(\omega, x, \mu_T) &=& \sum_{\text{initial pols.}} \langle p_1 | (\bar{\xi}_q W)_{n}(x) \>\frac{\bnslash}{2}\delta(\bar{{\cal P}}_n -\omega)(W^\dagger \xi_q)_{n}(0)| p_1\rangle, \nn \\
J_\bn^{\bar{q}}(\omega_2, y,\mu_T) &=& \sum_{\text{initial pols.}} \langle p_2 | \text{Tr}_{\text{spin}} \Big [  \>\frac{\nslash}{2}(W^\dagger \xi_q)_{\bn}(y)\delta(-\omega-\bar{{\cal P}}_\bn )(\bar{\xi}_q W)_{\bn}(0) \Big ]| p_2\rangle, \nn \\
S_{qq}(z,\mu_T) &=& \text{Tr} \langle 0 | \bar{T} [ S_n^\dagger S_\bn ] (z) \> T[ S_\bn^\dagger S_n ](0) |0\rangle.
\eea
Next we perform the integrals over $x^+$ and $y^-$ components in Eq.~(\ref{cross-section-3}) by defining the Fourier transformed jet functions as
\bea
\int \frac{dx^+}{4\pi} e^{\frac{i}{2}k_n^-x^+} J_n^{q}(\omega_1,x^+,x^-,x_\perp,\mu) &=& J_n^{q}(\omega_1,k_n^-,x^-,x_\perp,\mu),\nn \\
\int \frac{dy^-}{4\pi} e^{\frac{i}{2}k_\bn^+y^-} J_n^{q}(\omega_2,y^+,y^-,y_\perp,\mu) &=& J_\bn^{\bar{q}}(\omega_2,y^+,k_\bn^+,y_\perp,\mu), \nn \\
\eea
combining the label and residual momenta as
\bea
\int d\omega_1 dk_n^- &\to& \int d\omega_1, \qquad \int d\omega_2 dk_\bn^+ \to \int d\omega_2, \nn \\
\eea
and absorbing the residual momenta $k_n^-,k_\bn^+$ into $\omega_1,\omega_2$ respectively to get
\bea
\frac{d^2\sigma}{du\> dt}&=& \sum_{qijKL}\frac{\pi F^{KL;q}}{4Q^2N_c^2} \int \frac{dq^+dq^-}{2}\int d^2k_\perp \>\delta (q^+q^- - \vec{k}_\perp^2 - M_Z^2)  \int d\omega_1 \int d\omega_2 \>H_Z^{KL;ijq}(\omega_1,\omega_2,\mu_Q;\mu_T)  \nn \\
&\times& \int \frac{dx^-d^2x_\perp}{(2\pi)^3}\int \frac{dy^+ d^2y_\perp}{(2\pi)^3}\frac{1}{2}\int \frac{dz^+dz^-d^2z_\perp}{(2\pi)^4}  \int \frac{db^+db^-}{2(2\pi)^2}\int \frac{d^2b_\perp}{(2\pi)^2} \frac{1}{2}\int dk_n^+ d^2k_n^\perp \nn \\
&\times& \frac{1}{2}\int dk_\bn^- d^2k_\bn^\perp \frac{1}{2}\int dk_{s}^+dk_{s}^-d^2k_{s}^\perp  e^{-i \vec{k}_{n\perp}\cdot (\vec{x}_\perp -\vec{b}_\perp)} e^{-i \vec{k}_{\bn\perp}\cdot (\vec{y}_\perp -\vec{b}_\perp)}e^{-i \vec{k}_{s\perp}\cdot (\vec{z}_\perp -\vec{b}_\perp)} \nn \\
&\times& e^{\frac{i}{2}k_n^+(x^--b^-)}e^{\frac{i}{2}k_\bn^-(y^+-b^+)}e^{\frac{i}{2}k_{s}^-(z^+-b^+)}e^{\frac{i}{2}k_{s}^+(z^--b^-)} e^{\frac{i}{2}(\omega_1 - q^-)b^+} e^{\frac{i}{2}(\omega_2 - q^+)b^-}e^{i\vec{b}_\perp \cdot \vec{k}_\perp}\nn \\
&\times& \delta \big [u - M_Z^2 +  Q \>q^-\big ] \delta \big [t - M_Z^2 +  Q \>q^+ \big ] J_n^{q}(\omega_1, x^-,x_\perp,\mu_T) J_\bn^{\bar{q}}(\omega_2, y^+,y_\perp,\mu_T)S_{qq}(z,\mu_T). \nn \\
\eea
Performing the integrals over the momenta $k_n^+,k_\bn^-, k_{n,\bn}^\perp$ and $k_s^\mu$ and the $x,y,z$ coordinates, we get
\bea
\frac{d^2\sigma}{du\> dt}&=& \sum_{qij}\frac{\pi }{4Q^2N_c^2} \int \frac{dq^+dq^-}{2}\int d^2k_\perp \> \delta (q^+q^- - \vec{k}_\perp^2 - M_Z^2) \delta \big [u - M_Z^2 +  Q \>q^-\big ] \delta \big [t - M_Z^2 +  Q \>q^+ \big ]  \nn \\
&\times& \int d\omega_1 \int d\omega_2 \>H_Z^{ijq}(\omega_1,\omega_2,\mu_Q;\mu_T)    \int \frac{db^+db^-}{2(2\pi)^2}\int \frac{d^2b_\perp}{(2\pi)^2}e^{\frac{i}{2}(\omega_1 - q^-)b^+} e^{\frac{i}{2}(\omega_2 - q^+)b^-}e^{i\vec{b}_\perp \cdot \vec{k}_\perp} \nn \\
&\times&J_n^{q}(\omega_1, b^-,b_\perp,\mu_T) J_\bn^{\bar{q}}(\omega_2, b^+,b_\perp,\mu_T) S_{qq}(b^+,b^-,b_\perp,\mu_T),
\eea
where from brevity we have defined 
\bea
\label{hard-2}
H_Z^{q}(\omega_1,\omega_2,\mu_Q;\mu_T)  &\equiv& \sum_{KLij} F^{KL;q}H_Z^{KL;ijq}(\omega_1,\omega_2,\mu_Q;\mu_T). 
\eea
This expression can be brought into the form
\bea
\frac{d^2\sigma}{du\> dt}&=&\sum_{q} \frac{\pi }{4Q^2N_c^2}  \int dq^+dq^-\int d^2k_\perp  \int \frac{d^2b_\perp}{(2\pi)^2}e^{i\vec{b}_\perp \cdot \vec{k}_\perp}  \delta \big [u - M_Z^2 +  Q \>q^-\big ] \delta \big [t - M_Z^2 +  Q \>q^+ \big ] \nn \\
&\times& \delta (q^+q^- - \vec{k}_\perp^2 - M_Z^2)  \int d\omega_1 \int d\omega_2 \> (4 \omega_1 \omega_2)\>H_Z^{q}(\omega_1,\omega_2,\mu_Q;\mu_T)\nn \\
&\times&\int dk_n^+ dk_\bn^- B_n^{q}(\omega_1, k_n^+,b_\perp,\mu_T) B_\bn^{\bar{q}}(\omega_2, k_\bn^- ,b_\perp,\mu_T){\cal S}_{qq}(\omega_1-q^--k_\bn^-,\omega_2-q^+-k_n^+,b_\perp,\mu_T),\nn \\
\eea
where we have made use of the  Fourier transformed functions defined as
\bea
 B_n^{q}(\omega_1, k_n^+,b_\perp,\mu) &=&\frac{1}{2\omega_1} \int \frac{db^-}{4\pi} e^{\frac{i}{2}k_n^+b^-} J_n^{q}(\omega_1, b^-,b_\perp,\mu), \nn \\
 B_\bn^{\bar{q}}(\omega_2, k_\bn^-,b_\perp,\mu) &=& \frac{1}{2\omega_2} \int \frac{db^+}{4\pi} e^{\frac{i}{2}k_\bn^-b^+} J_\bn^{\bar{q}}(\omega_2, b^+,b_\perp,\mu), \nn \\
 {\cal S}_{qq}(\tilde{\omega}_1,\tilde{\omega}_2,b_\perp,\mu) &=& \int \frac{db^+db^-}{16\pi^2} e^{\frac{i}{2}\tilde{\omega}_1b^+} e^{\frac{i}{2}\tilde{\omega}_2b^-} S_{qq}(b^+,b^-,b_\perp,\mu).
\eea
The $B_{n,\bn}^{q,\bar{q}}$ functions are referred to as the purely collinear impact-parameter Beam Functions (iBFs) and are defined with a zero-bin subtraction to avoid double counting soft emissions already encoded in the soft function ${\cal S}_{qq}$. It was shown in \cite{Mantry:2009qz} that convolution over the purely collinear iBFs and the soft function can be written as
\bea
\frac{d^2\sigma}{du\> dt}&=& \sum_{q}\frac{\pi }{4Q^2N_c^2}  \int dq^+dq^-\int d^2k_\perp  \int \frac{d^2b_\perp}{(2\pi)^2}e^{i\vec{b}_\perp \cdot \vec{k}_\perp} \delta \big [u - M_Z^2 +  Q \>q^-\big ] \delta \big [t - M_Z^2 +  Q \>q^+ \big ]\nn \\
&\times& \delta (q^+q^- - \vec{k}_\perp^2 - M_Z^2) \int d\omega_1 \int d\omega_2 \>(4\omega_1\omega_2) H_Z^{q}(\omega_1,\omega_2,\mu_Q;\mu_T) \>\nn \\
&\times&\int dk_n^+ dk_\bn^- \tilde{B}_n^{q}(\omega_1, k_n^+,b_\perp) \tilde{B}_\bn^{\bar{q}}(\omega_2, k_\bn^- ,b_\perp){\cal S}^{-1}_{qq}(\omega_1-q^--k_\bn^-,\omega_2-q^+-k_n^+,b_\perp),\nn \\
\eea
where $\tilde{B}_{n,\bn}^{q,\bar{q}}$ are the `naive' iBFs  or simply the iBFs defined without a soft zero-bin subtraction, and ${\cal S}^{-1}_{qq}$ is the inverse Soft Function (iSF). Next we rewrite the cross section in terms the variables $x_1,x_2,t_n^+,t_\bn^-$ defined as
\bea
x_1= \frac{\omega_1}{Q}, \qquad x_2= \frac{\omega_2}{Q}, \qquad t_n^+ = x_1 Q k_n^+, \qquad t_\bn^- = x_2 Q k_\bn^-,
\eea
to get
\bea
\frac{d^2\sigma}{du\> dt}&=&\frac{\pi }{N_c^2}  \int dq^+dq^-\int d^2k_\perp  \int \frac{d^2b_\perp}{(2\pi)^2}e^{i\vec{b}_\perp \cdot \vec{k}_\perp}\> \delta \big [u - M_Z^2 +  Q \>q^-\big ] \delta \big [t - M_Z^2 +  Q \>q^+ \big ] \nn \\
&\times&\delta (q^+q^- - \vec{k}_\perp^2 - M_Z^2)  \int_0^1 dx_1\int_0^1 dx_2\>  \sum_{q} H_Z^{q}(x_1x_2Q^2,\mu_Q;\mu_T) \nn \\
&\times&\int dt_n^+ dt_\bn^- \tilde{B}_n^{q}(x_1, t_n^+,b_\perp) \tilde{B}_\bn^{\bar{q}}(x_2, t_\bn^- ,b_\perp){\cal S}^{-1}_{qq}(x_1Q-q^--\frac{t_\bn^-}{x_2 Q},x_2Q-q^+-\frac{t_n^+}{x_1Q},b_\perp),\nn \\
\eea
where we used the fact that $H_Z^{q}(\omega_1,\omega_2,\mu_Q;\mu_T) = H_Z^{q}(\omega_1 \omega_2, \mu_Q;\mu_T)$.
In the next step, the iBFs are matched onto the PDFs as 
\bea
\tilde{B}_{n,\bn}^{q}(x, t,b_\perp,\mu) &=& \int _x^1\> \frac{dz}{z}\> {\cal I}_{n,\bn; qr} (\frac{x}{z}, t,\mu) \>f_{r}(z,\mu),
\eea
so that the differential cross-section becomes
\bea
\frac{d^2\sigma}{du\> dt}&=& \frac{\pi^2 }{Q^2N_c^2}   \int_0^1 dx_1\int_0^1 dx_2 \int_{x_1}^1 \frac{dx_1'}{x_1'} \int_{x_2}^1 \frac{dx_2'}{x_2'}   \nn \\
&\times& \sum_{q}H_Z^{q}(x_1x_2Q^2,\mu) \>{\cal G}^{qrs}(x_1,x_2,x_1',x_2',u,t,\mu_T) f_r(x_1',\mu_T)  f_s(x_2',\mu_T),\nn \\  
\eea
with a sum over repeated indices understood. The function ${\cal G}^{qrs}$ is given by
\bea
 {\cal G}^{qrs}(x_1,x_2,x_1',x_2',u,t,\mu_T)&=&  \int \frac{d^2b_\perp}{(2\pi)^2} J_0\Bigg [b_\perp \sqrt{\frac{(M_Z^2-u)(M_Z^2-t)}{Q^2}-M_Z^2}\>\Bigg ]\>\nn \\
&\times&\int dt_n^+ dt_\bn^- \> {\cal I}_{n;qr}(\frac{x_1}{x_1'}, t_n^+,b_\perp,\mu_T)\>{\cal I}_{\bn;\bar{q}s}(\frac{x_2}{x_2'}, t_\bn^-,b_\perp,\mu_T)\nn \\
&\times& {\cal S}^{-1}_{qq}(x_1Q-\frac{(M_Z^2-u)}{Q}-\frac{t_\bn^-}{x_2 Q},x_2Q-\frac{M_Z^2-t}{Q}-\frac{t_n^+}{x_1Q},b_\perp).\nn \\
\eea
Using Eqs.(\ref{mandel-1}) and (\ref{mandel-2}) we can obtain the differential cross-section in terms of the $p_T$ and $Y$ variables 
\bea
\frac{d^2\sigma}{dp_T^2\> dY}&=& \frac{\pi^2 }{N_c^2}   \int_0^1 dx_1 \int_0^1 dx_2\int_{x_1}^1 \frac{dx_1'}{x_1'} \int_{x_2}^1 \frac{dx_2'}{x_2'}   \nn \\
&\times&   \sum_{q} H_Z^{q}(x_1x_2Q^2,\mu_Q;\mu_T) \>{\cal G}^{qrs}(x_1,x_2,x_1',x_2',p_T,Y,\mu_T) f_r(x_1',\mu_T)  f_s(x_2',\mu_T),\nn \\  
\eea
where
\bea
 &&{\cal G}^{qrs}(x_1,x_2,x_1',x_2',p_T,Y,\mu_T)=  \int \frac{d^2b_\perp}{(2\pi)^2} J_0\big [b_\perp p_T\big ]\>\int dt_n^+ dt_\bn^- \> {\cal I}_{n;qr}(\frac{x_1}{x_1'}, t_n^+,b_\perp,\mu_T)\>{\cal I}_{\bn;\bar{q}s}(\frac{x_2}{x_2'}, t_\bn^-,b_\perp,\mu_T)\nn \\
&\times& {\cal S}^{-1}_{qq}(x_1 Q-e^{Y}\sqrt{\text{p}_T^2+M_Z^2}-\frac{t_\bn^-}{x_2 Q}, x_2 Q-e^{-Y}\sqrt{\text{p}_T^2+M_Z^2}- \frac{t_n^+}{x_1 Q},b_\perp,\mu_T).\nn \\
\eea

\bibliographystyle{h-physrev3.bst}
\bibliography{DrellYan}

\end{document}